\documentclass[fleqn,usenatbib]{mnras}

\usepackage{newtxtext,newtxmath}

\usepackage[T1]{fontenc}

\DeclareRobustCommand{\VAN}[3]{#2}
\let\VANthebibliography\thebibliography
\def\thebibliography{\DeclareRobustCommand{\VAN}[3]{##3}\VANthebibliography}

\usepackage{graphicx}	
\usepackage{amsmath}	

\usepackage{orcidlink}
\usepackage{academicons}
\definecolor{orcidlogocol}{HTML}{A6CE39}

\usepackage{xcolor}

\newcommand{\starbug}{\textsc{starbugii}}
\newcommand{\jwst}{\textit{JWST}\ }
\newcommand{\hst}{\textit{HST}\ }
\newcommand{\euclid}{\textit{Euclid}\ }
\newcommand{\spitzer}{\textit{Spitzer}\ }
\newcommand{\Hi}{\mbox{\rm H{\small I}}}
\newcommand{\Hii}{\mbox{\rm H{\small II}}}

\graphicspath{{figures}}

\title[AGB Dust in NGC~6822]{JWST NIRCam and MIRI Reveal the Dust‑Producing AGB Population of NGC~6822}

\author[Nally et al.]{Conor Nally$^{1}$\thanks{E-mail: conor.nally@ed.ac.uk }\orcidlink{0000-0002-7512-1662},
Olivia C.\ Jones$^{2}$\orcidlink{0000-0003-4870-5547},
Laura Lenki\'{c}$^{3}$\orcidlink{0000-0003-4023-867},
Annette M.\ N.\ Ferguson$^{1}$\orcidlink{0000-0001-7934-1278},
Nolan Habel$^{4}$\orcidlink{0000-0002-2667-1676},
\newauthor{
Alec S.\ Hirschauer$^{5}$\orcidlink{0000-0002-2954-8622},
Margaret Meixner$^{4}$\orcidlink{0000-0002-0522-3743},
P.\ J.\ Kavanagh$^{6}$\orcidlink{0000-0001-6872-2358},
Martha L.\ Boyer$^{7}$\orcidlink{0000-0003-4850-9589},
Omnarayani Nayak$^{8}$\orcidlink{0000-0001-6576-6339},
}
\newauthor{
B.\ Sargent$^{9}$\orcidlink{0000-0001-9855-8261},
P.\ Scicluna$^{10,11}$\orcidlink{0000-0002-1161-3756}
}\\
\\
$^{1}$ Institute for Astronomy, University of Edinburgh, Blackford Hill, Edinburgh, EH9 3HJ, UK \\
$^{2}$ UK Astronomy Technology Centre, Royal Observatory, Blackford Hill, Edinburgh, EH9 3HJ, UK \\
$^{3}$ IPAC, California Institute of Technology, 1200 E. California Blvd., Pasadena, CA 91125, USA \\
$^{4}$ Jet Propulsion Laboratory, California Institute of Technology, 4800 Oak Grove Dr., Pasadena, CA 91109, USA \\
$^{5}$ Department of Physics \& Engineering Physics, Morgan State University, 1700 East Cold Spring Lane, Baltimore, MD 21251, USA \\
$^{6}$ Department of Physics, Maynooth University, Maynooth, Co. Kildare, Ireland \\
$^{7}$ Space Telescope Science Institute, 3700 San Martin Drive, Baltimore, MD 21218, USA \\
$^{8}$ NASA Goddard Space Flight Center, 8800 Greenbelt Road, Greenbelt, MD, USA \\
$^{9}$ SETI Institute, Mountain View, CA, 94043, USA \\
$^{10}$ Centre for Astrophysics Research, Department of Physics, Astronomy and Mathematics, College Lane Campus, University of Hertfordshire, Hatfield AL10 9AB, UK \\
$^{11}$ Space Science Institute, 4750 Walnut Street, Suite 205, Boulder, CO 80301, USA \\
}
\date{Accepted XXX. Received YYY; in original form ZZZ}

\pubyear{2026}

\begin{document}
\label{firstpage}
\pagerange{\pageref{firstpage}--\pageref{lastpage}}
\maketitle

\begin{abstract}
We present a photometric catalogue of the Local Group dwarf galaxy NGC~6822 based on deep {\em JWST} observations obtained with the Near‑Infrared Camera (NIRCam) and Mid‑Infrared Instrument (MIRI). Point‑spread‑function photometry and band‑matching were performed with \starbug. The resulting catalogue contains 864{,}114 NIRCam point sources and 17{,}235 MIRI detections, with 10{,}079 detected in both instruments.
Blackbody fitting yields effective temperatures and bolometric luminosities for 119{,}621 stars, providing a detailed view of the resolved stellar content.
Candidate evolved stars were selected from NIRCam–MIRI colour–magnitude diagrams, with the sample refined by removing resolved contaminants and excluding young stellar objects through spectral energy distribution (SED) fitting.  The final sample of 1226 evolved stars was analysed using the Grid of Red Supergiant and Asymptotic Giant Branch Models  (\textsc{grams}), from which dust‑production rates were obtained and carbon‑rich or oxygen‑rich asymptotic giant branch (AGB) classifications assigned via a likelihood‑weighted comparison across the full model set.
Across the {\em JWST} fields, evolved stars return $5.6\times10^{-7}\,M_\odot\,{\rm yr^{-1}}$ of dust to the interstellar medium,  with oxygen‑rich AGB stars contributing 60\% and carbon‑rich AGB stars 35\%, despite the low metallicity of NGC~6822. 
The unexpectedly high oxygen‑rich contribution indicates the presence of intermediate‑mass AGB stars undergoing hot‑bottom burning, in line with the recent star‑formation history of NGC~6822.
Dust‑producing AGB stars exhibit a centrally concentrated carbon‑rich population and a more extended oxygen‑rich population. We also identify {\em JWST} colour relations that provide robust photometric estimators of dust‑production rates for evolved stars. 
\end{abstract}

\begin{keywords}
galaxies: dwarf -- galaxies: irregular -- galaxies: individual (NGC~6822) -- infrared: galaxies -- infrared: stars -- stars: AGB and post-AGB
\end{keywords}



\section{Introduction}\label{sec:intro}

Thermally pulsing asymptotic giant branch (AGB) stars represent the final stage of evolution for low- to intermediate-mass stars (1--8~M$_\odot$; \citealp{Herwig2005}). They are key producers of dust in galaxies \citep{Hoppe2022}, enriching the interstellar medium (ISM). During this evolutionary phase, they can lose up to ${\sim}80\%$ of their initial mass through dust-driven mass loss \citep{Gall2011}. Although AGB stars evolve more slowly than high-mass stars, which end their evolution in supernova explosions, they can produce large quantities of dust in as little as 30~Myr \citep{Boyer2017}. Additionally, unlike supernovae—which may destroy dust produced during earlier evolutionary stages—AGB-produced dust is not subject to imminent destruction by the progenitor star, establishing AGB stars as crucial contributors to the massive dust reservoirs seen in high-redshift galaxies (e.g.,~$z > 6$; \citealt{Algera2023, Schneider2024a}).

Due to the complex physical processes that arise once a star begins to thermally pulse, understanding the subsequent evolution of AGB stars is challenging: their complexity means that they cannot be easily modelled, and observational constraints are limited, especially in low-metallicity environments. Uncertainties in stellar evolution models primarily stem from two processes: the third dredge-up (3DU), which enriches the stellar surface with carbon, and mass loss, both of which are strongly influenced by a star’s metallicity.

AGB stars separate into two chemical classes according to their surface carbon-to-oxygen ratio: carbon-rich (C-type; C/O $> 1$) and oxygen-rich (M-type; C/O $< 1$). The surface chemistry determines which molecules and dust species form in the stellar atmosphere, influencing opacity, dust-production efficiency, and subsequent evolution. As a result, C- and M-type AGB stars evolve at different rates, display distinct colours and luminosities, and imprint their environments in different ways. Because the formation of C-type AGB stars is favoured at low metallicity \citep{Karakas2002}, the ratio of C- to M-type stars provides a useful proxy for a galaxy’s metallicity \citep{Iben1983, Battinelli2005}.

The AGB populations of the Large and Small Magellanic Clouds (LMC, SMC), with metallicities of 0.5~Z$_\odot$ and 0.2~Z$_\odot$, respectively \citep{Russell1992}, have been studied extensively by the SAGE \emph{Spitzer} surveys \citep{Meixner2006, Gordon2011, Kemper2010}. Carbon stars generally exhibit redder colours than O-rich M-type stars, and although the chemistry of the reddest “extreme’’ AGB candidates could not be determined photometrically \citep{Blum2006}, spectroscopy has shown that these sources are predominantly carbon-rich \citep{Zijlstra2006, Sloan2016, Jones2017b}. In both the LMC and SMC, the number of O-rich AGB stars exceeds that of C-rich stars \citep{Blum2006, Boyer2011, Riebel2012}; however, the total dust input to the ISM is dominated by carbonaceous material \citep{Groenewegen2009}. Estimated AGB dust-return rates are $\dot{M} \sim 10^{-4}$~M$_\odot$\,yr$^{-1}$ in the LMC \citep{Srinivasan2009} and $\dot{M} \sim 10^{-6}$~M$_\odot$\,yr$^{-1}$ in the SMC \citep{Srinivasan2016}, with a small number of highly evolved “extreme’’ stars dominating the dust budget \citep{Boyer2011}.

Beyond the Magellanic Clouds, the DUSTiNGS survey extended mid-infrared AGB studies to 50 dwarf galaxies \citep{Boyer2015a}, finding no strong dependence of dust production on metallicity \citep{Boyer2015b}. This suggested that AGB stars may remain efficient dust producers even in very low-metallicity environments.

With {\em JWST} we can now observe the resolved stellar populations of galaxies in greater detail than ever before. For instance, \emph{JWST} observations of the Magellanic Clouds reach 10 magnitudes deeper than {\em Spitzer} and 2 magnitudes deeper than {\em Hubble} \citep{Jones2023a, Habel2024, Nayak2024a, Nayak2024b}, enabling the first extragalactic detection of brown dwarfs \citep{Zeidler2024}. In galaxies as distant as 18~Mpc, \emph{JWST} can resolve the AGB population down to the tip of the red giant branch (TRGB) \citep{Hirschauer2024} and thereby constrain their star formation histories (SFH) \citep{Bortolini2024}.

Given these capabilities, nearby dwarf galaxies such as NGC~6822 provide ideal laboratories for studying resolved stellar populations with {\em JWST}.
NGC~6822 is a Local Group ($d = 490$~kpc; \citealt{Fusco2012}), metal-poor irregular dwarf galaxy. Its evolved stellar populations have been extensively studied in the optical and near-infrared using ground-based observations \citep{Letarte2002, Cioni2005, Kang2006, Groenewegen2009, Battinelli2011, Sibbons2012, Kacharov2012, Sibbons2015, Whitelock2013}, providing constraints on the AGB population and C/M ratio, which is typically found to be $\sim0.2$--$0.6$ in the central regions.
NGC~6822 is generally believed to have evolved in tidal isolation \citep{Battinelli2006, Mcconnachie2021}, although some studies suggest a possible close encounter with the Milky Way around ${\sim}3$--4~Gyr ago \citep{Teyssier2012, Zhang2021}. The presence or absence of such an interaction remains an open question and has implications for the galaxy’s SFH. NGC~6822 has formed stars continuously for ${\sim}12$--15~Gyr \citep{Gallart1996c, Wyder2001}, with an increase in star formation rate (SFR) around ${\sim}2$--3~Gyr ago \citep{Tolstoy2001}. Current star formation is concentrated in highly localised \Hii{} regions, including the bright {\em Hubble} regions \citep{Cannon2006} and the more deeply embedded {\em Spitzer} regions \citep{Jones2019, Hirschauer2020, Lenkic2024}.

NGC~6822 exhibits an irregular morphology dominated by a central stellar bar. Its \Hi{} distribution is complex, extending into two prominent tails reaching more than $0.5^\circ$ from the galaxy’s centre, and includes a large $14\arcmin \times 10\arcmin$ cavity to the east \citep{deBlok2000}. The gas dynamics show significant structure \citep[and references therein]{Park2022}. The young ($\lesssim 180$~Myr) stellar population closely traces the \Hi{} distribution \citep{Komiyama2003, deBlok2006}, whereas the intermediate-age to old ($\gtrsim 2$~Gyr) population extends much further and may exhibit an elongated elliptical morphology almost perpendicular to the \Hi{} structure \citep{deBlok2006, Zhang2021}. These two components are likely unrelated, supporting the hypothesis that an interaction with a proposed companion dwarf to the north-west may have warped the \Hi{} disk and triggered star formation. Recent Euclid Early Release Observations have further demonstrated the galaxy’s complex stellar structure.  Euclid resolves individual stars and star clusters across NGC~6822’s main body and halo with its wide‑field near‑IR imaging, revealing detailed radial colour profiles, extended cluster populations, and multiple candidate globular clusters \citep{Hunt2025, Howell2026}.

\begin{table} 
	{\centering
    \caption{Summary of key NGC~6822 properties relevant to this study.}
	\begin{tabular}{lc} 
        \hline
        \multicolumn{2}{c}{Properties of NGC~6822}\\
        \hline
        Distance & 490 $\pm$ 40 kpc \citep{Mateo1998}\\
        Distance modulus $(m-M)_0$ & 23.45\\
        E(B-V) & 0.35 $\pm$ 0.04 \citep{Tantalo2022} \\
        Metallicity & [Fe/H]$=-1.05$~\citep{Kirby2013}\\
        \hline
	\end{tabular}
	\label{tab:characteristics}
    }
\end{table}

In this work, we combine the {\em JWST} survey of \citet{nally2024} with new data reduction and analysis to construct a full near‑ and mid‑infrared band–matched point-source catalogue of NGC~6822, and to constrain the properties and dust-production rates of its AGB star candidates. Section~\ref{sec:obs} describes the {\em JWST} observations, our data reduction pipeline, the point-spread-function (PSF) photometry, and the construction of the NGC~6822 catalogues. Section~\ref{sec:method} outlines the stellar-parameter fitting routines and the chemical classification of AGB stars. We present our results in Section~\ref{sec:res}, and summarise our conclusions in Section~\ref{sec:conclusions}.

\section{Observations and Data Reduction}\label{sec:obs}

\subsection{JWST Observations and Data Processing}

Imaging of the central stellar bar of NGC6822 with the {\em JWST} Near‑Infrared Camera (NIRCam; \citealt{Rieke2005, Rieke2023}) and Mid‑Infrared Instrument (MIRI; \citealt{Rieke2015, Wright2023}) was obtained as part of the GTO programme ID~1234 (PI: M.~Meixner) and was first presented in \citet{nally2024} and \citet{Lenkic2024}.

The NIRCam observations cover a $\sim 4\arcmin \times 6\arcmin$ region centred at $\mathrm{RA}=19{:}44{:}56.1990$, $\mathrm{Dec}=-14{:}47{:}51.29$, using four wide-band filters: F115W, F200W, F356W, and F444W. The $2\times1$ mosaic employed a primary \textsc{fullbox 4tight} dither pattern together with a three‑point subpixel dither sequence to optimally sample the point-spread function (PSF). The data were reduced using the {\em JWST} pipeline \citep[version 1.9.6;][]{bushouse_howard_2023_7714020} with CRDS context \texttt{jwst\_1075.pmap}.

The MIRI observations consist of a narrower $6\times1$ mosaic centred at $\mathrm{RA}=19{:}44{:}58.0949$, $\mathrm{Dec}=-14{:}48{:}20.620$, covering a $\sim 2\arcmin \times 7\arcmin$ region within the NIRCam footprint. Imaging was obtained in the F770W, F1000W, F1500W, and F2100W filters, using a four‑point \textsc{cycling} dither pattern and the \textsc{fastr1} readout mode. The MIRI data were reduced using {\em JWST} pipeline version 1.9.5 with CRDS version 11.16.21 and context \texttt{jwst\_1084.pmap}. During Stage~2 processing, a small astrometric correction was applied to align the MIRI frames to the GAIA‑DR3–matched NIRCam reference frame, ensuring high‑precision astrometry across the 1.15--21~$\mu$m wavelength range.

A complete description of the NIRCam and MIRI observations and processing steps is provided in \citet{nally2024}.

\subsection{Source Detection and Photometry}

\subsubsection{NIRCam PSF Photometry and Completeness}
\label{sec:photometry_nircam}

To construct a high‑reliability, band‑matched NIRCam– and MIRI–photometric catalogue of NGC\,6822, we adopt source positions, fluxes, and associated uncertainties directly from the NIRCam point‑source catalogue presented by \citet{nally2024}. We do not perform any additional data processing, photometry, or quality filtering on the NIRCam data beyond that described by \citet{nally2024}. Source detection and PSF photometry were performed on the reduced, 1/f‑noise–corrected, and \textit{Gaia}~DR3–aligned Stage~2 NIRCam exposures using the \starbug~PSF‑fitting package \citep{starbug2}. Only reliable point sources were retained; extended objects, cosmic rays, and other spurious artefacts were removed during catalogue construction. 
We adopt the completeness limits reported by \citet{nally2024}  (see their Table~2); the catalogue is complete to approximately $F115W \simeq 25.6$~mag and $F200W \simeq 24.6$~mag (Vega), with correspondingly brighter limits at longer wavelengths. The tip of the red giant branch occurs at $F200W \sim 17.5$, placing the evolved stellar populations analysed in this work several magnitudes above the completeness limit.

\subsubsection{MIRI PSF Photometry}
\label{sec:photometry_miri}

The NGC~6822 MIRI data exhibit complex and spatially variable background emission, making PSF photometry challenging. In addition, at the time of the earlier analyses, the available MIRI model PSFs did not fully characterise optical distortion effects \citep{Gaspar2021, Dicken2024}, further limiting the reliability of PSF‑based measurements. For these reasons, the catalogues in \citet{nally2024} and \citet{Lenkic2024} employed PSF photometry for the NIRCam observations but relied on aperture photometry for the MIRI data.

As a consequence, the MIRI aperture‑photometry catalogues for NGC~6822 exhibit larger flux uncertainties—particularly in regions with structured diffuse emission—and may include slightly extended sources such as faint background galaxies in low‑surface‑brightness areas or compact dusty enhancements in brighter regions.

Here, we conduct PSF photometry for the first time on the MIRI F770W, F1000W, F1500W, and F2100W data using \starbug~\citep{starbug2} version \textit{v0.7.3}, producing a high‑fidelity MIRI point‑source catalogue for NGC~6822 and improving the flux estimates and associated uncertainties of the extracted objects.

We perform source detection on the Stage‑3 single‑band mosaics, which provide the best signal‑to‑noise ratio and hence reveal the faintest dust‑embedded sources. Source detection is carried out using \texttt{starbug2 --detect} with the parameters listed in Table~\ref{tab:sb-miri}. These settings were optimised to identify low‑contrast sources even in regions with high and spatially varying backgrounds. Additional diagnostic parameters, such as {\sc sharp} and {\sc round}, provide quality assessments of each detection; their thresholds were chosen to exclude objects with values characteristic of cosmic rays, extended or resolved sources, clusters, blends, and other imaging artefacts.

A particular challenge in the MIRI data, relative to NIRCam, is the abundance of false detections arising from bright dust structures in the mid‑IR. To mitigate this, we introduce the \textsc{smoothness} parameter into \starbug. This quantity, defined as the ratio of the mean flux in the photometric aperture to that in a second aperture 1.5 times larger, measures how smoothly the surface brightness rises toward the source. It therefore provides an effective discriminator between genuine point sources and compact peaks in the diffuse ISM. Combined, these stringent detection and quality‑selection criteria ensure robust point‑source lists for each photometric band.

\begin{figure*}
    \centering
    \includegraphics[width=\textwidth]{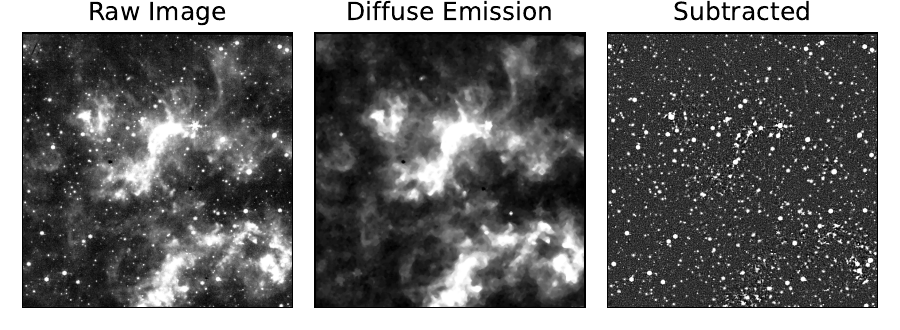}
    \caption{Diffuse-emission estimation for a single Stage‑2 exposure of the dusty Spitzer-I region of NGC~6822. The left panel shows the raw input array, the middle panel shows the result of \texttt{starbug2 --background}, and the right panel shows the background‑subtracted image on which PSF fitting is performed.}
    \label{fig:bgd}
\end{figure*}

Following detection on the Stage‑3 mosaics, point‑source photometry is performed on the individual Stage‑2 exposures, which provide more accurate fluxes than the mosaicked data. We generate 5\arcsec\ model PSFs for each of the four MIRI bands using {\sc webpsf} \citep{Perrin2014} version \textit{1.2.1}. Diffuse ISM emission is removed from the Stage‑2 exposures using \texttt{starbug2 --background}, and PSF fitting is then performed on the background‑subtracted images with \texttt{starbug2 --psf}. An example exposure, its estimated diffuse background, and the final background‑subtracted image are shown in Figure~\ref{fig:bgd}. This procedure requires a clean and complete input source list to derive an accurate diffuse‑emission model; any bright residuals typically correspond to detections that failed the geometric quality checks listed in Table~\ref{tab:sb-miri}.

Finally, all photometric measurements from the individual exposures in a given band are spatially matched using \texttt{starbug2-match} with the {\sc match\_thresh} values listed in Table~\ref{tab:sb-miri}. A source must be detected in at least three individual exposures to be retained in the final catalogue for that photometric band. This requirement guards against residual artefacts near frame edges or from cosmic‑ray contamination, and results in high‑fidelity point‑source catalogues for all four MIRI bands.

\starbug\ returns AB magnitudes for each detected source. These are transformed to the Vega system using the AB--to--Vega offsets from the CRDS reference files \texttt{jwst\allowbreak\_nircam\allowbreak\_abvegaoffset\allowbreak\_0002.asdf} and \texttt{jwst\allowbreak\_miri\allowbreak\_abvegaoffset\allowbreak\_0001.asdf} (see Table~\ref{tab:psf-details}).  The photometric uncertainties returned by \starbug\ are derived from the covariance matrix of the weighted PSF fit and therefore represent formal statistical uncertainties; as is standard for PSF‑fitting photometry, these may be underestimated for sources where additional systematics (e.g. PSF mismatch or correlated noise) are not fully captured.

\begin{table}
    \centering
    \caption{\starbug\ parameters used in the MIRI point‑source photometry, applied with \textsc{starbug2}~v0.7.3. A detailed description of each parameter can be found in the \starbug\ \href{https://starbug2.readthedocs.io/en/latest/}{documentation.}}
    \begin{tabular}{l|cccc}
    \hline
     Parameter& F770W & F1000W & F1500W & F2100W \\
    \hline
    {\sc sigsky}     & 1.8 & 1.8 & 1.8 & 1.8 \\
    {\sc sigsrc}     & 5.0 & 5.0 & 4.0 & 3.0 \\
    {\sc sharp\_lo}  & 0.4 & 0.4 & 0.225 & 0.2 \\
    {\sc sharp\_hi}  & 0.76 & 0.9 & 0.53 & 0.84 \\
    {\sc round1\_hi} & 0.9 & 1.0 & 0.7 & 0.5 \\
    {\sc round2\_hi} & 0.6 & 1.0 & 0.6 & 0.6 \\
    {\sc smooth\_hi} & 0.98 & 1.0 & 0.98 & -- \\
    {\sc ricker\_r}  & 2.0 & 2.0 & 5.0 & 5.0 \\
    \hline
    {\sc bgd\_size}  & 2 & 10 & 2 & 15 \\
    {\sc crit\_sep}  & 8 & 8 & 8 & 8 \\
    {\sc max\_xydev} & 3p & 3p & 3p & 3p \\
    \hline
    {\sc match\_thresh} & 0.15 & 0.2 & 0.2 & 0.25 \\
    \hline
    \end{tabular}
    \label{tab:sb-miri}
\end{table}

\begin{figure}
    \centering
    \includegraphics[width=\columnwidth]{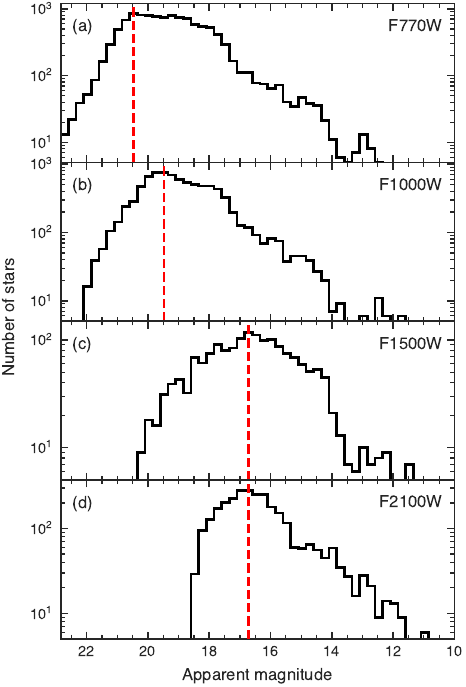}
    \caption[MIRI Luminosity functions]{Luminosity functions of the MIRI point‑source catalogues in the F770W, F1000W, F1500W, and F2100W bands. Vertical red dashed lines indicate the adopted completeness limits.}
    \label{fig:Lum_functions_MIRI}
\end{figure}

To assess the depth of the resulting catalogues, Figure~\ref{fig:Lum_functions_MIRI} shows the apparent‑magnitude luminosity functions of the MIRI point‑source catalogues in the F770W, F1000W, F1500W, and F2100W bands. The source counts rise toward fainter magnitudes before turning over and subsequently declining. We adopt this turnover magnitude as an empirical completeness limit for each band; the corresponding limits are listed in Table~\ref{tab:psf-details}.

\subsubsection{Cross-Band and Cross-Instrument Matching}

The four single‑band MIRI point‑source catalogues were merged into a single combined MIRI catalogue with \texttt{starbug2-match $--$band}. The matching algorithm uses a simple nearest‑match criterion and allows sources detected with high reliability in some, but not all, bands to be included in the combined MIRI point‑source catalogue. As progressively longer-wavelength catalogues with larger PSF FWHM values are incorporated, we adopt increasing match-separation thresholds (Table~\ref{tab:psf-details}). Taking F770W as the base catalogue, we match F1000W with a threshold of 0.15\arcsec, followed by F1500W and F2100W with thresholds of 0.2\arcsec\ and 0.25\arcsec, respectively. Sources without a positional match are appended to the catalogue, with their coordinates taken from the shortest-wavelength band in which they are detected. This approach preserves astrometric accuracy while accounting for the differing photometric depths of the MIRI filters.

The F770W catalogue provides the shortest-wavelength MIRI band with which to connect the MIRI and NIRCam catalogues. To combine these catalogues, we require that each MIRI F770W source be matched to a NIRCam source detected in F444W in the \citet{nally2024} catalogue. We expect that most F770W detections will also be present in F444W, as typical stellar and warm-dust spectral energy distributions (SEDs) are brighter at $4.44\,\micron$ than at $7.7\,\micron$. If a closer positional match exists in the NIRCam data but that object lacks an F444W detection, the match is rejected. If a source is not detected in F770W, the next-longest available MIRI band is used for the F444W cross-match. This criterion ensures that the NIRCam/MIRI cross-match is made between bands tracing the same underlying sources and avoids spurious matches arising from the much higher source density in the NIRCam catalogue. We adopt a $0.3\arcsec$ separation threshold for the F770W--F444W match to minimise mismatches across the wide wavelength baseline. However, the PSF FWHM increases by more than an order of magnitude from the shortest NIRCam band (F115W) to the longest MIRI band (F2100W). Consequently, compact groups of NIRCam sources may appear blended into single MIRI detections at longer wavelengths. In such cases, the full MIRI flux is assigned to the nearest NIRCam source with an F444W detection in the cross‑match, even though the measured MIRI flux may originate from several NIRCam‑resolved sources within the MIRI PSF. 

\setlength{\tabcolsep}{4pt}
\begin{table}
    \centering
    \caption[Filter properties]{Filter properties, source counts, and magnitude limits for the NGC\,6822 NIRCam and MIRI catalogue.}
    \begin{tabular}{lcccccc}
    \hline
    \hline
    Filter & Number & Pixel & FWHM & AB--Vega & Completeness \\
    Name   & Sources& Scale & [arcsec] & Zeropoint & (Vega mag) \\
    \hline
    F115W  & 739\,604 & 0.031\arcsec & 0.040 & 0.764 & 25.60 \\
    F200W  & 538\,946 & 0.031\arcsec & 0.066 & 1.686 & 24.58 \\
    F356W  & 167\,800 & 0.063\arcsec & 0.116 & 2.811 & 23.21 \\
    F444W  & 142\,005 & 0.063\arcsec & 0.145 & 3.238 & 23.09 \\
    \hline
    F770W  & 11\,258 & 0.11\arcsec & 0.269 & 4.384 & 20.47 \\
    F1000W & 8\,257  & 0.11\arcsec & 0.328 & 4.956 & 19.47 \\
    F1500W & 1\,616  & 0.11\arcsec & 0.488 & 5.839 & 16.72 \\
    F2100W & 2\,261  & 0.11\arcsec & 0.674 & 6.532 & 16.72 \\
    \hline
    \end{tabular}
    \label{tab:psf-details}
\end{table}

\section{Results and Analysis}\label{sec:method}

\subsection{Description of Point Source Catalogue}

We present the band‑matched NGC~6822 PSF–photometry catalogue constructed from NIRCam and MIRI observations. The resulting catalogue contains 864{,}114 NIRCam point sources and 17{,}235 MIRI detections, with 10{,}079 detected in both instruments.
Table~\ref{tab:cat_description}  summarises the catalogue contents and column definitions. 

For each source, we provide ICRS right ascension (RA) and declination (Dec), measured from the shortest‑wavelength band in which the source is reliably detected to ensure the highest astrometric precision. Observed (i.e. uncorrected for extinction) Vega magnitudes are reported for all available NIRCam and MIRI filters, with corresponding photometric uncertainties listed in columns prefixed with “e” (e.g. F770W and eF770W);  these uncertainties are derived from the PSF-fitting implemented in \starbug\ \citep{Nally2026}.  In cases where multiple NIRCam detections fall within the MIRI PSF, the full MIRI flux is assigned to the nearest NIRCam source with an F444W detection. The catalogue also provides effective temperatures (T\textsubscript{eff}) and bolometric luminosities (L\textsubscript{bol}) derived as described in Section~\ref{sec:method:bb}. The complete catalogue will be available in machine‑readable form through VizieR and as supplementary material accompanying this paper.

\begin{table}
    \caption[Description of revised point source catalogue]{Description of point source catalogue columns. 
    All NULL values in floating-point columns are represented as ``nan''; other columns do not contain NULL values. }
    \begin{tabular}{c|l}
         \hline
    \hline
         Column Name & Description \\
         \hline
         Catalogue$\_$Number & The identifying name of the source \\
         RA  & Right ascension (ICRS; deg) \\
         Dec & Declination (ICRS; deg) \\
         Teff & Effective temperature from blackbody fitting (K)\\
         Lbol & Bolometric luminosity from blackbody fitting (L$_\odot$)\\
         \hline
         F115W  & Magnitude in F115W NIRcam filter (Vegamags)\\
         F200W  & Magnitude in F200W NIRcam filter (Vegamags)\\
         F356W  & Magnitude in F356W NIRcam filter (Vegamags)\\
         F444W  & Magnitude in F444W NIRcam filter (Vegamags)\\
         eF115W & Error on F115W magnitude \\
         eF200W & Error on F200W magnitude \\
         eF356W & Error on F356W magnitude \\
         eF444W & Error on F444W magnitude \\
         \hline
         F770W  & Magnitude in F770W MIRI filter (Vegamags)\\
         F1000W  & Magnitude in F1000W MIRI filter (Vegamags)\\
         F1500W  & Magnitude in F1500W MIRI filter (Vegamags)\\
         F2100W  & Magnitude in F2100W MIRI filter (Vegamags)\\
         eF770W & Error on F770W magnitude \\
         eF1000W & Error on F1000W magnitude \\
         eF1500W & Error on F1500W magnitude \\
         eF2100W & Error on F2100W magnitude \\
        \hline
    \end{tabular}
    \label{tab:cat_description}
\end{table}

\subsection{Foreground/Background Removal and Extinction Corrections}
\label{sec:contamination}

Before performing the analysis described in the following sections, we apply a series of cleaning and photometric corrections to the catalogue. We correct the photometry for foreground reddening using the \texttt{dust\_extinction} package~\citep{Gordon2024}. We adopt the value E(B$-$V)\,$=0.35$~\citep{Tantalo2022} to correct for the moderate Galactic foreground extinction and apply the extinction curve of \cite{Cardelli1989} assuming $R_V=3.1$. This corresponds to $A_{\rm F115W}\sim0.4$\,mag and $\lesssim0.05$\,mag in the MIRI bands; consequently, small variations in the foreground extinction are unlikely to significantly affect our results.

Foreground stars were removed by cross-matching the catalogue with \emph{Gaia} Data Release~3 \citep{GAIAeDR3} using the procedure outlined in \citet{nally2024}. No additional foreground stars were identified in the new MIRI PSF catalogue for NGC~6822, consistent with the substantially brighter detection limits of \emph{Gaia} relative to the MIRI observations.
Contamination from resolved background galaxies is largely suppressed by the \starbug\ detection and cleaning routines. Any remaining extended or non-stellar sources are removed by visual inspection as described in Section~\ref{sec:method:selection}.

\subsection{Blackbody Fitting}\label{sec:method:bb}

To obtain approximate temperatures and luminosities for the sources in our combined NGC~6822 NIRCam and MIRI catalogue, we fit each SED with a blackbody model (Eq.~\ref{equ:bb}), 
\begin{equation}
    \rm F_\nu(A,T_{eff})=A\frac{2h\nu^3}{c^2}\frac{1}{e^{\frac{h\nu}{k_bT_{eff} }}-1}
    \label{equ:bb}
\end{equation}
where $\rm F_\nu(A,T_{eff})$ is the flux at frequency $\rm \nu$ for a star with effective temperature $\rm T_{eff}$. Scale factor \textit{A} scales the model linearly to account for distance, the speed of light $c$, the Boltzmann constant $k_B$ and the Planck constant $h$.

The parameters $A$ and $T_{\rm eff}$ are varied freely to minimise the $\chi^2$ value using \texttt{scipy.optimize.minimize}, computed between the observed photometric points and the model fluxes in linear space. 
To ensure a well‑constrained fit, we require each source to be detected in at least four of the eight photometric bands. This criterion yields 119{,}621 sources in the \jwst\ catalogue with valid temperature estimates.
The bolometric luminosity $L_{\rm bol}$ is then calculated by integrating the blackbody function using the fitted values of $A$ and $T_{\rm eff}$. The resulting effective temperatures and bolometric luminosities are included in the NGC~6822 point‑source catalogue (Table~\ref{tab:cat_description}). These correspond primarily to the brighter sources in our catalogue, with temperature estimates approximately complete to F200W~$\sim$22.5~mag, around $\sim$0.5~mag below the red clump indicating the limiting magnitude at which reliable blackbody fits can be obtained.

\begin{figure}
    \centering
    \includegraphics[width=\columnwidth]{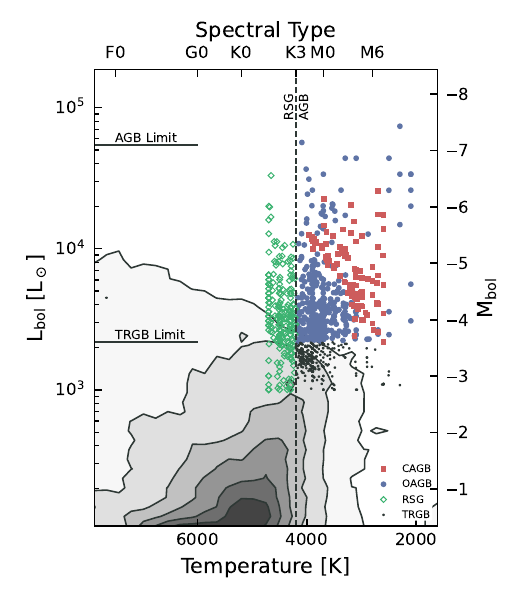}
    \caption[H–R diagram of the central bar of NGC~6822]{Hertzsprung–Russell (H–R) diagram of the central bar of NGC~6822.  The $x$‑axis shows effective temperature ($T_{\rm eff}$) and the $y$‑axis shows bolometric luminosity ($L_{\rm bol}$) in solar units on the left and the corresponding bolometric magnitude on the right. 
    The underlying greyscale density map displays the blackbody temperatures and luminosities derived for all fitted point sources in the NGC~6822 catalogue (Section~\ref{sec:method:bb}), with darker colours indicating higher source density. 
    Blue circles mark oxygen‑rich AGB (O‑AGB) stars and red squares represent carbon‑rich AGB (C‑AGB) stars.
    Horizontal lines at $M_{\rm bol} = -3.6$ and $M_{\rm bol} = -7.1$ indicate the classical TRGB and AGB limits, respectively, and the dashed vertical line at 4200\,K (K3\,III) marks the division between AGB and RSG stars.
    }
    \label{fig:hr-diag}
\end{figure}

The H–R diagram for the 119{,}621 fitted sources, shown as a greyscale density map in Figure~\ref{fig:hr-diag}, exhibits a well‑defined RGB. AGB stars, identified in Section~\ref{sec:method:grams}, populate the region above the TRGB, consistent with expectations for their evolutionary phase and chemistry.

There are two important caveats to obtaining well‑defined estimates of effective temperature and luminosity using blackbody fitting.
First, hot (blue) stars have blackbody peaks at wavelengths shorter than those sampled by the NIRCam filters. The peak of $\rm F_\nu$ in Equation~\ref{equ:bb} occurs at a wavelength $\lambda$ satisfying $\lambda T = 5099.3,\mu{\rm mK}$ \citep{Zhang2012}. Consequently, any source with $T_{\rm eff} \gtrsim 4500~{\rm K}$ will have its SED peak in the UV or optical—outside the \jwst\ wavelength coverage—leading to increasingly poorly constrained fits at higher temperatures, such that large $T_\mathrm{eff}$ values primarily reflect model degeneracy rather than precise physical temperatures. The most reliable $T_{\rm eff}$ values in our NGC~6822 sample therefore correspond to K‑ and M‑type stars, whose SEDs peak in the near‑infrared.

Second, sources exhibiting infrared excess—such as AGB stars with circumstellar dust or young stellar objects (YSOs) with circumstellar disks or envelopes—depart significantly from a pure blackbody SED. In such cases, the integrated blackbody fit overestimates the bolometric luminosity.  More physically motivated luminosity estimates for dusty AGB stars are derived in Section~\ref{sec:method:grams}, and YSO modelling is discussed in Section~\ref{sec:res:clean} and \citet{Lenkic2024}.

\subsection{Selecting Evolved Star Candidates}\label{sec:method:selection}

\begin{figure}
    {
    \centering
    \includegraphics[]{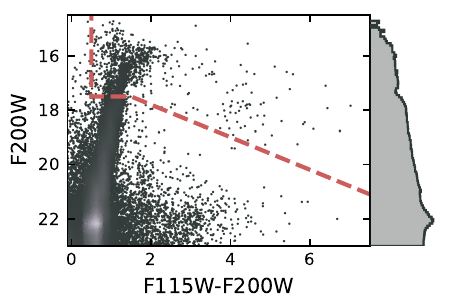}
    \includegraphics[]{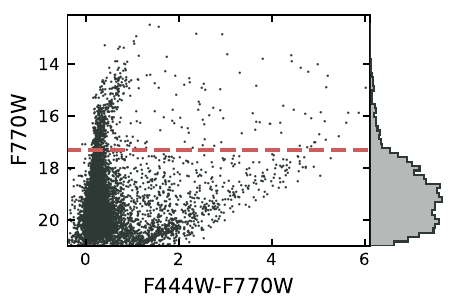}
    \includegraphics[]{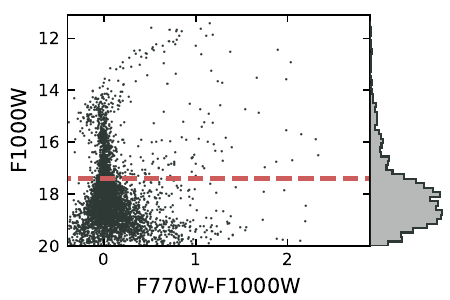}
    \includegraphics[]{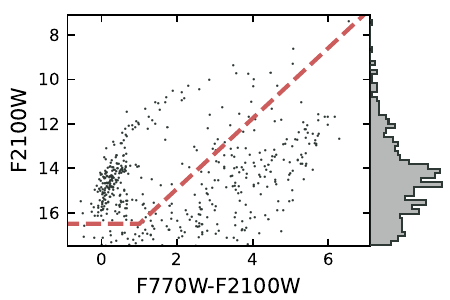}
    }
	\caption[CMD selection criteria for evolved stars]{CMDs of NGC~6822 showing the evolved-star selection boundaries. Four CMDs are presented: F200W versus F115W--F200W, F770W versus F444W--F770W, F1000W versus F770W--F1000W, and F2100W versus F770W--F2100W.  Selection boundaries are shown in red, with their numerical values listed in  Table~\ref{tab:agbselection}.     
    Magnitude distributions are shown in the histograms to the right of each CMD; the top CMD uses a logarithmic scale due to the large number of sources, while the remaining histograms use linear scaling.}
    \label{fig:method:sel}
\end{figure}

Building on the temperature and luminosity estimates derived in Section~\ref{sec:method:bb},  we next identify evolved‑star candidates within the NGC~6822 point‑source catalogue.  Dusty evolved stars are not well described by simple blackbody models,  so we apply colour–magnitude selections to construct an inclusive candidate list.

The TRGB provides a practical luminosity threshold for these cuts.  The bolometric TRGB magnitude for NGC~6822, computed using the metallicity‑dependent  relation of \citet{Salaris2005} and adopting [Fe/H] = –1.05 from \citet{Kirby2013}, is 
$M_{\rm bol}^{\rm TRGB} = -3.61$. This is consistent with the empirical TRGB magnitudes adopted by \citet{nally2024} for the colour–magnitude selections. Guided by these TRGB measurements and the source classifications from \citet{nally2024},  we define a series of NIRCam and MIRI colour–magnitude boundaries, shown in  Figure~\ref{fig:method:sel} and listed in Table~\ref{tab:agbselection}. A point source is included in our provisional evolved‑star candidate list if it satisfies at least one of these criteria.

The F115W--F200W versus F200W CMD forms the basis of the NIRCam selection. This colour index spans the wavelength range that brackets the blackbody peak of K–M–type evolved stars and is therefore sensitive to modest variations in effective temperature. We therefore adopt a horizontal cut at the TRGB (F200W~$\sim$17.5) and a sloped boundary from (1.5, 17.5) to (9.0, 22.0) to include evolved stars reddened by circumstellar dust whose extinction shifts them below the TRGB at redder F115W--F200W colours.
The F444W--F770W versus F770W and F770W--F1000W versus F1000W CMDs extend the selection into the mid‑IR, where dust emission becomes progressively more significant. In both CMDs, we place the boundary slightly below the TRGB  (F770W~$\sim$17.3; F1000W~$\sim$17.4) and retain all sources brighter than these limits. 
In the F770W--F2100W versus F2100W CMD, a diagonal boundary is adopted that follows the empirical separation between AGB stars and YSOs established by  \citet{Jones2017a}. This criterion enables the inclusion of heavily dust‑enshrouded evolved stars undergoing a superwind while mitigating contamination from star‑forming sources.

\begin{table}
    \centering
    \caption[Evolved star colour-cut selection]{Colour--magnitude selection criteria for evolved stars. For each colour--magnitude combination, the boundary coordinates are listed as (colour, magnitude) pairs. Sources lying above these boundaries are included in the provisional evolved--star candidate list.}
    \begin{tabular}{c|c}
         \hline
    \hline
         Colour      & Boundary \\
         Combination & Coordinates \\
         \hline
         F115W--F200W vs F200W   & (0.5,17.5) (1.5,17.5) (9.0,22.0) \\
         F444W--F770W vs F770W   & (-0.5,17.3) (7.0,17.3)  \\
         F770W--F1000W vs F1000W & (-0.5,17.4) (3.0,17.4)  \\
         F770W--F2100W vs F2100W & (-1.0,16.5) (1.0,16.5) (7.0,7.0) \\
         \hline
    \end{tabular}
    \label{tab:agbselection}
\end{table}

These cuts are deliberately broad and are not intended to produce a clean AGB sample, but are instead designed to ensure high completeness of the evolved‑star population, even at the expense of admitting sources that occupy similar regions of colour–magnitude space.  Such contaminants include dust‑accreting YSOs, unresolved star‑forming galaxies, and post‑AGB stars \citep{Jones2017b}.

In total, 1456 sources in the \jwst\ NGC~6822 point-source catalogue have at least four photometric flux measurements and fulfil at least one of the evolved star colour cuts. These sources constitute the provisional NGC~6822 evolved star candidate list.

\begin{figure*}
    \centering
    \includegraphics[width=\textwidth]{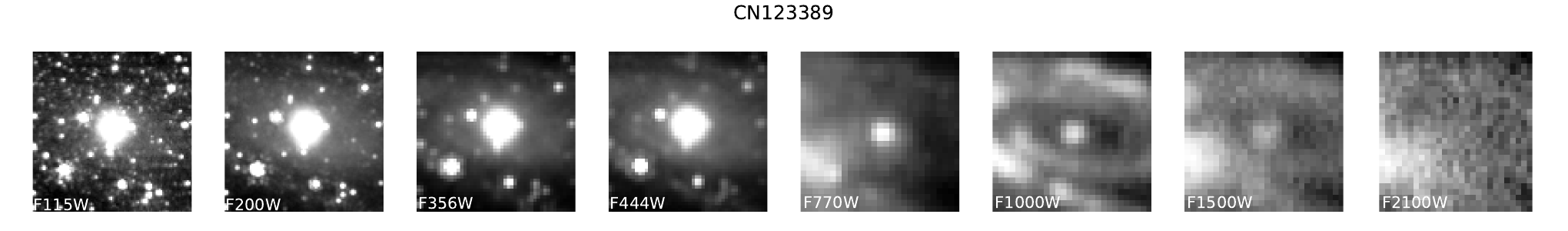}
    \includegraphics[width=\textwidth]{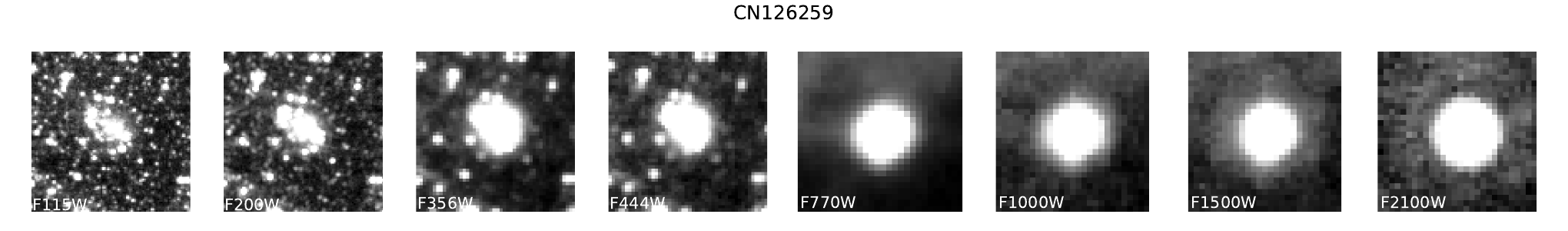}
    \caption[Resolved contaminants in the source list]{Zoom-in images of two contaminating sources in the provisional evolved star candidate list,shown across the eight \jwst\ mosaics, increasing in wavelength from left (F115W, 1.15$\mu$m) to right (F2100W, 21$\mu$m). The upper row shows source CN123389, a spiral galaxy with a point-like core. The lower row shows source CN126259, which appears as a point source in the MIRI images but is resolved in the NIRCam data, revealing a compact stellar cluster.}
    \label{fig:res:inspect}
\end{figure*}

Using the high spatial resolution of the NIRCam images, we visually inspect the candidate evolved stars to identify any resolved background galaxies that may remain in the catalogue. 
Resolved background galaxies with compact, point-like nuclei can appear morphologically similar to dust-embedded stellar sources in the \starbug\ detection routine. For example, the upper panel of Figure~\ref{fig:res:inspect} shows a galaxy with a compact nucleus, which has resolved structure due to its elongated disc and dust lanes that are visible at specific wavelengths.

In addition to these background galaxies, compact stellar clusters may also mimic single evolved stars in the longer-wavelength images. 
The spatial resolution of the MIRI images is $\gtrsim$0.638 pc at the distance of NGC~6822, while NIRCam images are $\lesssim$0.344 pc.
As a result, the broader MIRI PSF can cause stars in compact clusters to blend into a single point‑like source at longer wavelengths.
The lower panel of Figure~\ref{fig:res:inspect} shows one such object, which appears point-like in all MIRI images, even exhibiting PSF wings. With the resolving power of NIRCam, we resolve this object into a compact stellar cluster and exclude it from the evolved‑star sample.

The provisional colour-selected candidate list contained 56 objects that fell into one of these two categories. These objects are removed from the final evolved‑star sample.

\subsection{Evolved Star SED Modelling}\label{sec:method:grams}

Thermally pulsing AGB stars are efficient producers of dust, with the most extreme sources undergoing a superwind and losing mass at a rate of $\sim10^{-4}\,M_\odot\,\mathrm{yr^{-1}}$~\citep{Vassiliadis1993}. The amount and composition of the dust formed are closely tied to the chemistry of the circumstellar envelope. Molecular and dust signatures in the spectra of the star can be traced with well-sampled SEDs \citep{Jones2017a}. Following a similar methodology to \citet{Riebel2012}, \citet{Srinivasan2016} and \citet{Jones2018}, we fit the SEDs of the NGC~6822 evolved star candidates with a grid of radiative‑transfer models for evolved stars to confirm their evolutionary status, chemically classify them into C- and O-rich stars, and measure their dust production rates.

The photometric data for each of the 1400 colour-selected evolved-star candidates in NGC~6822 are fitted to the 80\,843 model SEDs in the Grid of Red Supergiant and Asymptotic Giant Branch ModelS (\textsc{grams}; \citealt{Srinivasan2011,Sargent2011}). The \textsc{grams} grid is divided into carbon-rich and oxygen-rich model sets, computed using the \textsc{2Dust} radiative-transfer code \citep{Ueta2003} together with stellar photosphere models appropriate for carbon-rich \citep{Aringer2009} and oxygen-rich \citep{Kucinskas2005,Kucinskas2006} stars. 
The O-rich dust is modelled using astronomical silicates \citep{Ossenkopf1992}. For C-rich stars, a combination of amorphous carbon (90\%) and silicon carbide (10\%) is used.

The \textsc{grams} models were computed for representative metal-poor stars $Z \sim 0.5\,Z_\odot$ in the Magellanic Clouds~\citep{Bernard2008}, comparable to that of NGC~6822.
The O-rich models of \citet{Sargent2011} span effective temperatures $T_{\rm eff} = 2100$--$4700$\,K, optical depths at $10\,\micron$ ($\tau_{10}$) from $10^{-4}$ to 26.0, and inner-shell radii $R_{\rm in} = 3$, 7, 11, and 15\,$R_{\rm star}$, producing models with bolometric luminosities $L_{\rm bol} = 10^3$--$10^6\,L_\odot$.
The C-rich grid generated by \citet{Srinivasan2011} spans $2600 \leq T_{\rm eff} \leq 4000$\,K, optical depths at $11.3\,\micron$ ($\tau_{11.3}$) from $10^{-3}$ to 4.0, and inner-shell radii $R_{\rm in} = 1.5$, 3, 4.5, 7, and 12\,$R_{\rm star}$, corresponding to bolometric luminosities $1100 \leq L_{\rm bol} \leq 26\,000\,L_\odot$. The C-rich models also include the parameter $C2O$, the stellar surface carbon-to-oxygen ratio, with values of 1.4, 2, and 5. 
For convenience, we designate the O-rich models with $C2O = 0$. This choice does not represent a physical abundance, but serves solely to distinguish them from the C-rich grid. In practice, stars entering the AGB phase typically have $C/O \approx 0.6$ \citep{Lattanzio2004}.
Dust mineralogy is fixed and not varied in the grid. These parameters provide the key constraints needed to determine the evolutionary stage and chemical type of the evolved-star candidates in NGC~6822.

Each NGC~6822 evolved-star candidate with at least four photometric data points is compared against every model SED ($i$), where $i$ indexes the models in the \textsc{grams} grid, to evaluate the fit between the observed photometry and the model predictions.  For the SED fitting, the model SEDs are convolved with the appropriate \jwst\ filter transmission curves, and the model flux densities $F_\nu$ are scaled by a factor $A$ to correct for the adopted distance to NGC~6822 (490\,kpc) relative to the LMC distance (50\,kpc) at which the models were originally computed.

Following \citet{Srinivasan2016}, we compute a reduced $\chi^{2}$ per photometric point (i.e.\ $\chi^{2}$ divided by the number of flux measurements) as the fit-quality metric for each model. This value is denoted $\chi^{2}_{i}$, where $i$ indexes the model SED, thereby producing a distribution of fit qualities for each source. Stars whose minimum $\chi^{2}_{i}$ (their best-fitting model) exceeds $10^{4}$ are flagged as poor fits to all \textsc{grams} models and are removed from the evolved-star sample.

Each model in the grid has a likelihood given by $\rm\mathcal{L}_i\propto w_i=e^{-0.5\chi_i^2}$, which, with flat priors, is proportional to its posterior probability.  These likelihoods are then used as weights when computing averaged model parameters, allowing better-fitting models to contribute more strongly while retaining information about the full range of plausible fits and the uncertainty in the underlying model choice.

Chemical discrimination in the \textsc{grams} fits arises from how strongly the likelihood distribution favours one model grid over the other. Sources with detections across both the NIRCam and MIRI filters provide strong constraints on circumstellar dust properties, while those detected only in the NIRCam bands are primarily sensitive to the photospheric SED shape and molecular features. 
When wavelength coverage is limited, the likelihood distributions over the \textsc{grams} model grid become broader, reducing the ability to distinguish between carbon-rich and oxygen-rich chemistries. 

The parameters of primary interest in this work are the effective temperature ($T_{\rm eff}$), bolometric luminosity ($L_{\rm bol}$), and dust-production rate (DPR). For each model parameter $x$ and each evolved-star candidate, we compute a weighted average of the \textsc{grams} model fits to obtain representative physical values. The weighted mean of parameter $x$ is calculated using Equation~\ref{equ:weightedmean}, with weights $w_i$ derived from the model likelihoods, and the corresponding weighted standard deviation is computed using Equation~\ref{equ:weightedstdv}, where $n$ is the number of models. This approach incorporates the full model grid, with each model contributing in proportion to its fit quality, and thus provides estimates with an effective resolution finer than the discrete sampling of the grid itself. It also mitigates the risk of solutions becoming trapped in local minima in the $\chi^{2}$ parameter space.

\begin{equation}
    \bar{x}=\frac{\sum w_i x_{i}}{\sum w_i}
    \label{equ:weightedmean}
\end{equation}

\begin{equation}
    \sigma=\sqrt{ \frac{\sum w_i(x_i-\bar{x})^2}{(n-1)\sum w_i}}
    \label{equ:weightedstdv}
\end{equation}

We note that uncertainties derived from the likelihood-weighted standard deviation can become artificially small when the likelihood distribution is strongly peaked, particularly for sources with poorly constrained dust emission. In such cases, the weighting scheme concentrates on a small number of models, leading to an underestimation of the true parameter uncertainties. This differs from the approach adopted in previous {\sc grams} studies \citep[e.g.,][]{Riebel2012}, which estimate uncertainties using the median absolute deviation of a fixed subset of best-fitting models.

\begin{figure*}
    \centering
    \includegraphics[width=0.49\textwidth, trim={0.5cm 0.5cm 0 0}]{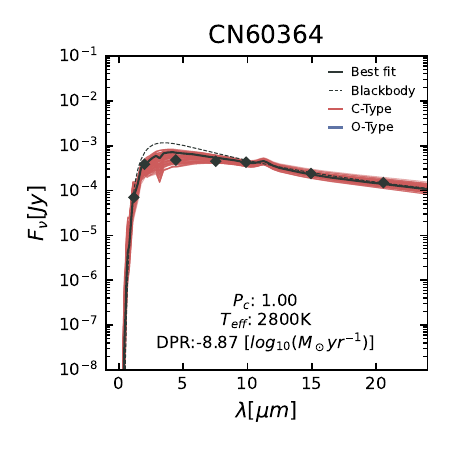}
    \includegraphics[width=0.49\textwidth, trim={0 0.5cm 0.5cm 0}]{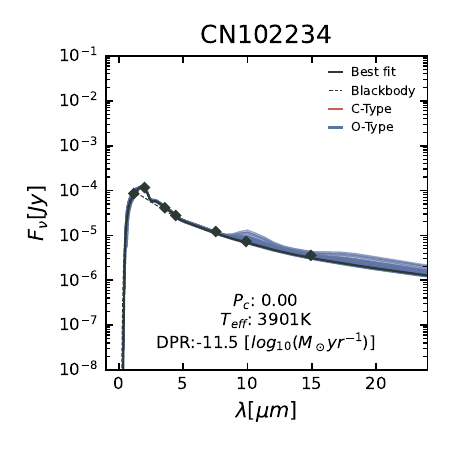}
    \includegraphics[width=0.49\textwidth, trim={0.5cm 0 0 0}]{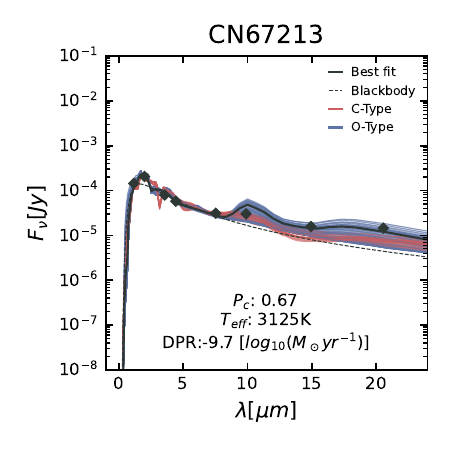}
    \includegraphics[width=0.49\textwidth, trim={0 0 0.5cm 0}]{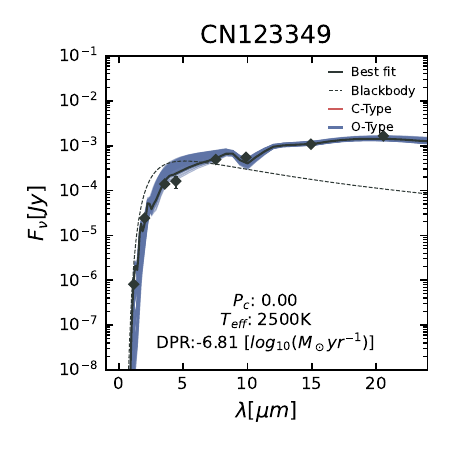}
    \caption[SED fitting with the \textsc{grams} models]{Four examples of \textsc{grams} SED fitting for evolved sources in NGC~6822. Each panel shows the top 100 best-fitting \textsc{grams} models to a given source (photometric points shown as black diamonds, with flux uncertainties where visible). Carbon-rich models are plotted in red and oxygen-rich models in blue. The solid black line indicates the best-fitting model, and the dashed line shows the best-fitting blackbody from Section~\ref{sec:method:bb}.}
    \label{fig:res:sed}
\end{figure*}

Figure~\ref{fig:res:sed} shows the SEDs and best-fitting \textsc{grams} models for four example sources. The figure panels display the photometric points as black diamonds, with lines indicating the 100 best-fitting models (for illustrative purposes): carbon-rich models are shown in red and oxygen-rich models in blue. The model with the lowest $\chi^{2}$ is plotted as a solid black line, and the dashed line represents the best blackbody fit.
The upper-left panel shows the dusty star CN60364, for which carbon-rich models provide good fits. The upper-right panel shows CN102234, a dust-free star. The lower-left panel displays CN67213, a moderately dusty source for which both carbon-rich and oxygen-rich models fit well. Finally, the lower-right panel shows CN123349, which exhibits a significant infrared excess and has the highest dust-production rate in the sample (see Section~\ref{sec:res:dpr}).

\begin{table}
    \centering
    \caption[Results of \textsc{grams} model fitting]{Description of the columns in the table containing the results of the \textsc{grams} model fits to the NGC~6822 evolved-star candidates.}
    \begin{tabular}{l|l}
        \hline
        \hline
        Column Name & Description \\
        \hline
        Catalogue\_Number & Unique source identifier \\
        $T_{\rm eff}$ & Effective temperature (K) \\
        $L_{\rm bol}$ & Bolometric luminosity ($\mathrm{L_\odot}$) \\
        DPR & Dust-production rate ($\mathrm{M_\odot\,yr^{-1}}$) \\
        $P_C$ & Probability of being a carbon star \\
        class$^a$ & Source classification \\
        \hline
        e$T_{\rm eff}$ & Standard deviation of $T_{\rm eff}$ \\
        e$L_{\rm bol}$ & Standard deviation of $L_{\rm bol}$ \\
        eDPR & Standard deviation of DPR \\
        \hline
    \end{tabular}

    $^a$ Defined in Section~\ref{sec:res:classification}.
    \label{tab:cat2_description}
\end{table}

Results from the \textsc{grams} model fitting are summarised in Table~\ref{tab:cat2_description}. The unique identifier for each source is listed in the column \texttt{Catalogue\_Number} and can be cross-referenced with Table~\ref{tab:cat_description}. The effective temperature ($T_{\rm eff}$), bolometric luminosity ($L_{\rm bol}$, in $\mathrm{L_\odot}$), and dust-production rate (DPR, in $\mathrm{M_\odot\,yr^{-1}}$), all derived from the weighted means of the model fits, are reported in separate columns. The $L_{\rm bol}$ and $T_{\rm eff}$ values obtained from the \textsc{grams} fitting supersede those from the blackbody fits and should be used when available. The corresponding weighted standard deviations are provided in the columns prefixed with `e'. The chemical classification, expressed as the probability of being a carbon star ($P_C$), along with the source classification (`class'), are discussed further in Section~\ref{sec:res:classification}. The evolved-star catalogue will be provided in electronic form via VizieR and as supplementary material accompanying this paper.

\subsection{Removing YSO contaminants} \label{sec:res:clean}

Evolved stars are not unique in infrared colour space, and are potentially confused with several contaminant populations, including YSOs, unresolved background galaxies, and planetary nebulae, all of which exhibit rising SEDs that remain bright in the mid-IR \citep{Jones2017a,Nayak2024a,Habel2024,Lenkic2024}. The colour-cut selections described in Section~\ref{sec:method:selection} are designed to limit the inclusion of such objects in the evolved-star candidate list; however, some contaminants inevitably remain.

Visual inspection reduces contamination from clearly resolved sources, as described in Section~\ref{sec:method:selection}, and the exclusion of objects that are poorly fitted by the \textsc{grams} models further limits non–evolved-star contaminants. However, removing contaminant YSOs is more challenging, as they appear point-like in the \jwst\ images. YSO SEDs often rise steeply in the mid‑IR, although extreme dust‑producing AGB stars can exhibit similar behaviour \citep{Jones2017a,Nayak2024a,Habel2024}. 
To identify potential YSO interlopers, we fit each source in the evolved star catalogue with the ``spubhmi'' model YSO SEDs from \citet{Richardson2024}. These are the model SEDs from \citet{Robitaille2017} convolved with the \jwst\ filter throughput functions. Similar to the {\sc grams} SED fits discussed above, the SED fitter provided by \citet{Robitaille2007} tests the observed photometry against every model in the ``spubhmi'' model grid, where the visual extinction $A_{V}$ is the single free parameter, and returns a $\chi^{2}$ per photometric point value for each as a goodness-of-fit metric \citep[$\chi^{2}_{\mathrm{YSO}}$; see][]{Robitaille2007}.  This method was applied by \citet{Lenkic2024} to the central star-forming region of NGC~6822; here, we extend it to the full \jwst\ footprint, but only for sources that satisfy the evolved-star colour cuts.

The results of the YSO and evolved-star fitting can be compared using the Akaike Information Criterion (AIC). This provides a principled means of balancing model fit quality against model complexity. The AIC is defined as: $\rm AIC=2k - 2ln\mathcal{L}_{max}=\chi^2+2k$, where $\rm\mathcal{L}_{max}$ is the maximised likelihood and $\rm k$ is the number of free parameters in the model. The first term rewards models that reproduce the data well (i.e.\ those with low $\chi^{2}$), while the second penalises models with additional free parameters. The AIC is therefore well-suited to comparing results obtained from different model-fitting approaches that may not use the same fitting methods. By placing the two fits on a common scale, the AIC allows us to determine which model family provides the better explanation of the data, rather than relying solely on the minimum $\chi^{2}$ value.

\begin{figure}
    \centering
    \includegraphics[width=\columnwidth]{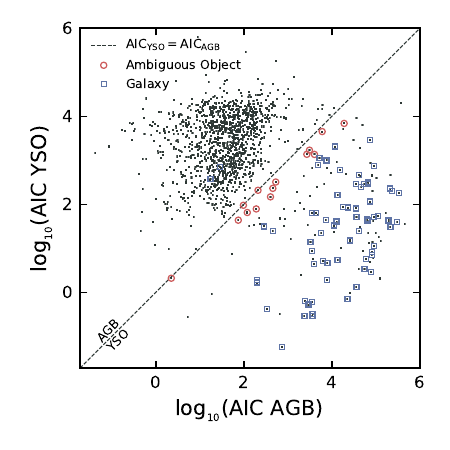}
    \caption[Removing YSO contaminants with a quality-of-fit comparison]{Comparison of the evolved-star and YSO model quality-of-fit values for each source.  Blue squares mark objects identified as resolved background galaxies.  The black dashed line indicates the ${\rm AIC_{YSO}} = {\rm AIC_{AGB}}$ relation; sources above this line exhibit SEDs more consistent with evolved stars than with YSOs.  Red circles highlight objects within $\Delta{\rm AIC} \leq 0.5$ of the boundary, which may correspond to heavily dust-enshrouded or variable evolved stars.  Spectroscopic observations will be required to confirm the true evolutionary nature of these sources.}
    \label{fig:res:cleaning}
\end{figure}

Figure~\ref{fig:res:cleaning} illustrates this comparison. Points located toward the left of the diagram indicate good evolved-star fits from the \textsc{grams} models, while points near the bottom indicate good YSO fits. The black dashed line marks the ${\rm AIC_{YSO}} = {\rm AIC_{AGB}}$ relation; sources lying above this line exhibit SEDs more consistent with evolved stars than with YSOs. We do not treat this boundary as absolute. Objects with ambiguous evolutionary status, defined as those with $\Delta{\rm AIC} \leq 0.5$, are retained as tentative evolved-star candidates, as they may represent extreme dust-producing AGB stars.
The blue squares indicate sources that show resolved structure during visual inspection of the \jwst\ images (Section~\ref{sec:method:selection}). Their positions on the diagram demonstrate the effectiveness of this approach in separating evolved stars from contaminating sources.

Of the 1400 sources that passed the colour cuts, 174 are rejected due to poor fits to the \textsc{grams} models (see Section~\ref{sec:method:grams}). Of the remaining 1226 sources, 1212 are best fitted by the evolved-star models, while a further 14 fall within $\Delta{\rm AIC}\leq 0.5$ of the dividing line.

\subsection{Classification of the Evolved-Star Sample}
\label{sec:res:classification}

Having established a clean evolved-star sample (Section~\ref{sec:res:clean}), we now classify these sources into their respective evolutionary and chemical types. We separate the evolved sources into four categories: oxygen- or carbon-rich AGB stars, red supergiants (RSGs), and stars fainter than the theoretical TRGB (hereafter simply TRGB). Their temperatures and luminosities, as derived from the \textsc{grams} models, are overlaid on Figure~\ref{fig:hr-diag} in green (RSG), blue (O-AGB), red (C-AGB), and black (TRGB). A summary of the classification counts is provided in Table~\ref{tab:classcount}.

The first separation is between RSG and AGB stars. The oxygen-rich \textsc{grams} models do not distinguish between O-AGB and RSG stars, so we apply a simple temperature cut at $T_{\rm eff}=4200$\,K to separate these populations. Stars hotter than this threshold correspond approximately to spectral type K3 or earlier, and we therefore classify them as RSGs, following \citet{Sibbons2012}, \citet{Hirschauer2020}, and \citet{Tantalo2022}. This boundary is shown as the dashed vertical line in Figure~\ref{fig:hr-diag}. In total, 253 stars are classified as RSGs, spanning the full luminosity range of the AGB population. Although RSGs with $M_{\rm bol}<-7.1$ are expected \citep{Wood1983} and have been observed in NGC~6822 \citep{Massey1998,Levesque2012,Patrick2015,Hirschauer2020,Antoniadis2025}, saturation in the NIRCam images reduces the number of these luminous RSGs that are recovered.

The TRGB marks the boundary between the RGB and AGB evolutionary phases. Adopting the bolometric TRGB magnitude derived in Section~\ref{sec:method:selection}, $M_{\rm bol}^{\rm TRGB} = -3.61$, we separate the AGB and RGB populations using a luminosity cut at this value in the H–R diagram. Early AGB stars and RGB stars have dust-free photospheres that can be fit by some models in the \textsc{grams} grid, as they have similar spectral types~\citep{Sargent2011}. We therefore classify sources cooler than 4200\,K and brighter than the TRGB as AGB stars, and those below this limit as TRGB sources. In total, 610 stars satisfy the AGB criteria, while 363 sources fall below the TRGB.

\begin{equation}
    \rm M^{\rm TRGB}_{\rm bol}=-0.19[Fe/H]-3.81
    \label{equ:trgb}
\end{equation}

Finally, the AGB stars are separated into two chemical classes: oxygen‑rich (O-AGB) and carbon‑rich (C-AGB). Unlike other parameters in the \textsc{grams} models, such as $T_{\rm eff}$ or DPR, the chemical classification is a binary property of the circumstellar envelope. When the best-fitting models include both chemistries, a probabilistic classification scheme is required.

We define $P_{C}$ as the probability that a given star is carbon-rich. Each \textsc{grams} model is assigned a value of 0 (oxygen-rich) or 1 (carbon-rich), and the weighted mean of these values—computed using Equation~\ref{equ:weightedmean} and the model likelihoods—yields $P_{C}$ in the range 0–1. Stars for which the highest-likelihood models are predominantly carbon-rich have $P_{C}$ close to unity, while those with mostly oxygen-rich fits have low $P_{C}$. Values of $P_C$ close to 0 or 1 indicate that the available data strongly favour one chemical type, while intermediate values arise when both carbon- and oxygen-rich models provide comparably good fits.  This typically occurs when the SED is weakly constrained by the available photometry, for example, when few data points are available. 

Adopting a threshold of $P_{C}=0.5$, we classify 191 AGB stars as C-AGB ($P_{C}>0.5$) and 419 as O-AGB ($P_{C}<0.5$). These populations are shown in Figure~\ref{fig:hr-diag}, where C-AGB stars are plotted in red and O-AGB stars in blue. 

Within this framework, the dependence of the classification on the number of photometric data points is naturally encoded. Sources with broader wavelength coverage, particularly those including MIRI photometry, typically yield sharply peaked likelihood distributions and hence robust classifications ($P_C \rightarrow 0$ or 1). In contrast, sources fitted with only four bands generally exhibit broader probability distributions, resulting in a higher fraction of intermediate $P_C$ values. Importantly, this reflects increased uncertainty rather than a systematic bias toward either chemical type.

Previous infrared surveys of evolved stars in Local Group galaxies \citep[e.g.][]{Srinivasan2009,Riebel2012,Jones2017a} generally adopted the single best-fitting SED to determine the chemical type of each AGB star. Our approach incorporates information from the full distribution of acceptable models and therefore differs from the best-fitting classification in a minority of cases. By comparing $P_{C}$ with the chemical type of the best-fitting model, we find that all stars we classify as oxygen-rich also have oxygen-rich best fits, whereas approximately 10\,per\,cent of stars classified as carbon-rich have an oxygen-rich best-fitting model. The remaining carbon-rich stars all have carbon-rich best-fitting models.

\begin{figure}
    \centering
    \includegraphics[width=\columnwidth]{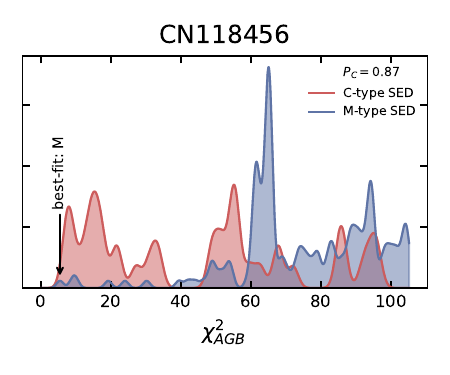}
    \caption[Distribution of $\chi^{2}$ values from SED fitting]{Distribution of $\chi^{2}$ values from fitting carbon-rich (red) and oxygen-rich (blue) \textsc{grams} model SEDs to a single source (CN118456). Of the top 70 best-fitting models (those with $\chi^{2}<40$), 90\,per\,cent are carbon-rich, yet the formal best-fitting \textsc{grams} model is oxygen-rich. Under the approach adopted by \citet{Srinivasan2009}, \citet{Riebel2012}, and \citet{Jones2017a}, this source would therefore be classified as oxygen-rich. However, our probabilistic method assigns it a carbon-rich probability of $P_{C}=0.87$, and we thus classify it as carbon-rich.}
    \label{fig:chi2dist}
\end{figure}

Figure~\ref{fig:chi2dist} illustrates one such case (CN118456). The distributions of $\chi^{2}$ values for carbon-rich (red) and oxygen-rich (blue) models show that the majority of good-fitting models favour a carbon-rich classification, yet the formal best-fitting model is oxygen-rich. This arises from a local minimum in the $\chi^{2}$ distribution associated with a single oxygen-rich model. By constructing the full probability density function from the weighted model ensemble, we obtain a more robust probabilistic classification: CN118456 has $P_{C}=0.87$ and is therefore most likely to be carbon-rich.

\subsubsection{Comparison with Previous AGB Classifications}

As an initial validation of our evolved-star selection, we compare our sample with the Mira and semi-regular variables identified by \cite{Battinelli2011} and \cite{Whitelock2013} in NGC~6822. Within our  {\em JWST} field, 23 of 32 variables from \cite{Whitelock2013} and all 13 from \cite{Battinelli2011} are matched to our evolved-star catalogue within 1\arcsec.

We next compare with the narrow-band catalogue of carbon stars from \cite{Letarte2002}. Within the {\em JWST} field, 87 optically identified carbon stars are matched to our evolved-star sample, of which 62 are classified as carbon-rich and 25 as oxygen-rich in our analysis.  While there is broad agreement, some optically identified carbon stars are instead classified as oxygen-rich, likely reflecting the higher spatial resolution of  {\em JWST}, which resolves sources blended in ground-based data.

Finally, we cross-match our sample with the spectroscopic AGB catalogue of \cite{Kacharov2012}. Within the {\em JWST} field, 27 sources are common to both samples, with 10 carbon-rich and 8 oxygen-rich sources classified consistently between the two datasets. One source (CN74628) is classified as a carbon star in our analysis but identified as an S0II star in the spectroscopic study. Such S-type stars represent an intermediate chemical class and are not explicitly captured by our binary carbon- and oxygen-rich classification scheme. 

Overall, these comparisons suggest that our evolved-star selection and classifications are broadly consistent with previous studies, with differences arising primarily from the higher spatial resolution of {\em JWST}, which resolves blended sources \citep[e.g.,][]{Lenkic2024}, its enhanced sensitivity to dust-enshrouded sources at longer infrared wavelengths, variations in wavelength coverage and spatial extent, and differing AGB selection criteria. Complete agreement with other studies is not expected given differing scientific aims and methodologies; rather, these complementary approaches together provide a more complete census of NGC~6822's stellar population.

\section{Discussion}\label{sec:res}

\subsection{Carbon Star Luminosity Function}

The luminosity function of C-rich stars provides a key insight into the third dredge-up of AGB stars, which is influenced by nucleosynthesis, surface chemical enrichment, and mass loss~\citep{Marigo1999}.
The carbon star luminosity function typically peaks around a bolometric magnitude of $M_{\rm bol}\sim -4.5$ to $-5.0$, a range set by the underlying core mass–luminosity relation for AGB evolution~\citep{Paczynski1970,Iben1983}, with variations produced by the progenitor mass and metallicity~\citep{Marigo2008}.
Third dredge-up efficiency in ${\sim}1.5$--$4\,M_\odot$ AGB stars determines the luminosity at which a star can become carbon-rich. High‑efficiency dredge‑up in low‑mass cores allows stars to become carbon‑rich earlier and at lower bolometric luminosities~\citep{Karakas2002}.
This is expected in lower metallicity environments, such as NGC~6822, as fewer third dredge-up events are required to raise the surface chemistry to C/O$>$1~\citep{Marigo1999,Groenewegen2006,Marigo2008}.
Alternatively, in higher‑mass AGB stars, the most luminous carbon‑rich stars may reach $\rm M_{\rm bol}{\sim}-6.0$ during late-stage superwind; however, this phase is short‑lived (${\sim}500$\,yr) and such stars are statistically rare~\citep{Vassiliadis1993,Hofner2018}. 
Observationally, the peak of the luminosity function reflects not only the physics of stellar interiors but also the cumulative effects of star-formation history, selection biases, and circumstellar dust obscuration, all of which influence which stars are detected and measured~\citep{Cioni2006b}.

\begin{figure}
    \centering 
    \includegraphics[width=\columnwidth]{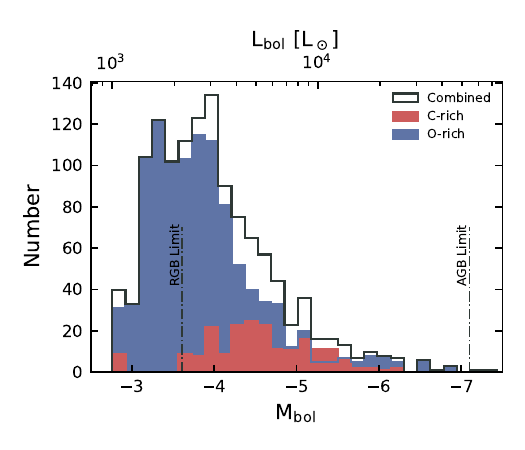}
    \caption[Bolometric luminosity function of evolved stars]{Luminosity function of the evolved stars in NGC~6822. The distribution of carbon-rich stars is shown in red. 
    All M-type sources (O-AGB, RSG, and TRGB stars) are shown in blue, and the cumulative distribution is overlaid in black. The classical RGB and AGB luminosity limits are indicated.}
    \label{fig:lf}
\end{figure}

The luminosity function of the evolved-star population of NGC~6822 is shown in Figure~\ref{fig:lf}, for both the C-rich and O-rich (O-AGB, RSG, TRGB) stars. 
Although C-rich AGB stars are rarely observed to exceed $M_{\rm bol} = -6.0$, theoretical models place the limit closer to $M_{\rm bol} = -6.4$~\citep{Iben1983}. 
Above this threshold, stars are massive enough (${\sim}4\,M_\odot$) to undergo Hot-bottom burning (HBB), where the convective envelope touches the hydrogen-burning shell~\citep{Boothroyd1992}. Temperatures at the base of the envelope reach ${\sim}3\times10^{7}\,{\rm K}$, preventing continued dredge-up of carbon and instead converting it to nitrogen through the CNO cycle~\citep{Ventura2005a}, thus setting an upper mass limit for C‑rich AGB stars. 
The brightest carbon‑rich star in the sample occurs at $M_{\rm bol} = -6.3$ ($L_{\rm bol} = 26\,000\,L_\odot$), and we find three C-AGB stars brighter than the observational ceiling of $M_{\rm bol} = -6.0$. 
The median C-AGB bolometric magnitude, $\bar{M}^{C}_{\rm bol} = -4.48 \pm 0.6$, is consistent with the value measured for C-rich Miras in NGC~6822 by \citet{Whitelock2013,Whitelock2013cat} ($M_{\rm bol} = -4.54$, assuming a distance modulus of 23.45). 
Similarly, \citet{Srinivasan2016} find that the carbon-star luminosity function in the SMC peaks at $M_{\rm bol} = -4.5$. The comparable metallicities of the two galaxies imply similar dredge‑up efficiencies, resulting in nearly identical luminosity-function peaks. 
This is despite their differing star-formation histories, which would influence the progenitor-mass distribution of AGB stars.  Both galaxies, however, experienced substantial star formation within the last ${\sim}1$--$6$\,Gyr~\citep{Harris2004,Fusco2014}, producing a large population of low- to intermediate-mass AGB progenitors visible today. 

In the full luminosity-function distribution, our sample contains two oxygen-rich sources brighter than the classical AGB limit of $M_{\rm bol} = -7.1$. This limit corresponds to the maximum luminosity achievable during steady hydrogen-shell burning on the thermally pulsing AGB, as set by the AGB core mass–luminosity relation \citep{Wagenhuber1998}, beyond which the degenerate helium core would ignite \citep{Iben1983,Vassiliadis1993}. 
These sources may instead be RSGs~\citep{Hirschauer2020,Antoniadis2025}, thermally pulsing AGB stars observed during a pulse that briefly increases their luminosity above the classical AGB limit \citep{Lattanzio2004}, or higher‑mass (8--10\,$M_\odot$) ``super‑AGBs'' that bridge the evolutionary gap between AGB stars and RSGs \citep{Siess2006}.

\subsection{Evolved Star Dust Production} \label{sec:res:dpr}

The luminosity and, consequently, the radiation pressure of an AGB star increases as it evolves through the thermally pulsing stage, allowing higher rates of mass loss to occur \citep{Vassiliadis1993,Bloeker1995}. Earlier in the evolution of AGB stars, the density of the stellar wind is too low for efficient dust-grain growth, resulting in negligible IR excess~\citep{Hofner2018}. As AGB stars evolve and their mass-loss rates increase, conditions become favourable for dust formation. Observations of AGB stars in the Magellanic Clouds have shown that this process produces a distinct dusty population above a threshold dust‑production rate (DPR) of $\mathrm{DPR \geq 10^{-11.3}\,}M_\odot\,\mathrm{yr^{-1}}$~\citep{Boyer2012,Riebel2012,Srinivasan2016}. Using the DPR values from the \textsc{grams} models, we find that NGC~6822 shows the same behaviour: a distinct population of dust‑producing AGB stars appears above a threshold of $\mathrm{DPR} \gtrsim 10^{-11.3}\,M_\odot\,yr^{-1}$, consistent with the Magellanic Cloud results.

\begin{figure}
	\centering
	\includegraphics[width=\columnwidth]{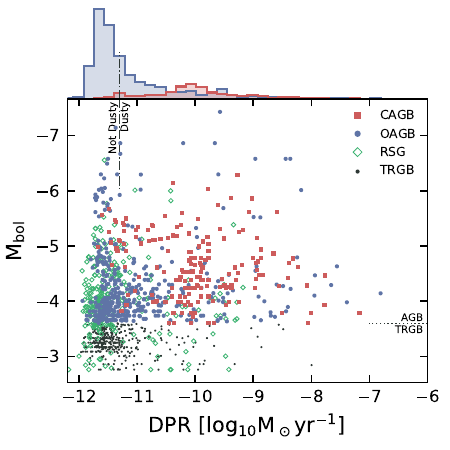}
	\caption[Dust-production rates of evolved stars]{Dust-production rate (DPR) versus bolometric magnitude ($M_{\rm bol}$). Oxygen-rich AGB stars are plotted in blue, carbon-rich AGB stars in red, RSG stars in green, and TRGB stars in black. The classical TRGB limit ($M_{\rm bol}=-3.61$) and the boundary between dust-free and dust-producing stars are overlaid. The upper panel shows the distribution of DPR values for the oxygen-rich dust produced by O-AGB and RSG stars (blue) and for the carbon-rich dust produced by C-AGB stars (red).}
	\label{fig:dpr}
\end{figure}

Figure~\ref{fig:dpr} shows the measured DPR (in $\log_{10}\,M_\odot\,\mathrm{yr^{-1}}$) versus $M_{\rm bol}$ for our NGC~6822 evolved-star sample. A histogram of binned DPR values for oxygen-rich (blue) and carbon-rich (red) dust is shown in the upper panel. We can divide Figure~\ref{fig:dpr} into two distinct populations: dust-free sources that extend vertically upward in bolometric magnitude, and dust-producing sources that have dust-production rates greater than $10^{-11.3}\,M_\odot\,\mathrm{yr^{-1}}$.

Table~\ref{tab:classcount} summarises the cumulative DPR from each class of evolved star. The total DPR from all evolved stars within the {\em JWST} field in the central bar of NGC~6822 is $5.6\times10^{-7}\,M_\odot\,\mathrm{yr^{-1}}$. AGB stars contribute 95\% of this total, with 4\% produced by sources below the TRGB brightness limit and the remaining 1\% by RSG stars. A small number of stars dominate the dust input: four sources account for 52\% of the total. Overall, 60\% of the dust production is oxygen-rich and 35\% carbon-rich, indicating that O-rich AGB stars dominate the dust return in NGC~6822.

\begin{table}
    \centering
    \caption[Dust production by evolved stars in NGC~6822]{Dust production by evolved stars in NGC~6822. For each stellar class, the total number of sources is listed, along with the number of objects with $\mathrm{DPR}\geq10^{-11.3}\,M_\odot\,\mathrm{yr^{-1}}$ and their cumulative DPR. AGB stars are shown both as a whole and separated into their carbon-rich and oxygen-rich subclasses.}
    \begin{tabular}{c|cccc}
        \hline
    \hline
        Stellar       & Number & Sources with           & Total DPR     & Percentage of\\
		Classification&        & $\mathrm{DPR}\geq10^{-11.3}$& [$\rm M_\odot\,yr^{-1}]$ & Total DPR\\
        \hline
        AGB  & 610 & 376 & $5.3\times10^{-7}$ & 95\\
        \hline
        C-AGB & 191 & 180 & $1.9\times10^{-7}$ & 35\\
        O-AGB & 419 & 196 & $3.4\times10^{-7}$ & 60\\
        RSG  & 253 & 61  & $7.5\times10^{-9}$ & 1\\
        TRGB & 363 & 106 & $2.4\times10^{-8}$ & 4\\
        \hline
        Total& 1226& 544 & $5.6\times10^{-7}$ & \\
        \hline
    \end{tabular}

    \label{tab:classcount}
\end{table}

Most C-AGB stars in the sample are efficient producers of dust, with 180/191 (94\%) producing dust at rates higher than $10^{-11.3}\,M_\odot\,\mathrm{yr^{-1}}$. These C-AGB stars have undergone at least one third dredge-up event, bringing carbon to the stellar surface.  The high opacity of the carbonaceous dust that forms in the circumstellar envelope allows efficient transfer of momentum from the stellar radiation to the wind. As such, the presence of carbon in an AGB photosphere leads to enhanced DPR~\citep{Lattanzio2004,Karakas2014}. 
The total quantity of carbonaceous dust produced by C-AGB stars in the central bar of NGC~6822 is $\sum \mathrm{DPR}_{C} = 1.9\times10^{-7}\,M_\odot\,\mathrm{yr^{-1}}$. The majority of this is produced by a single source with $\mathrm{DPR}=6.8\times10^{-8}\,M_\odot\,\mathrm{yr^{-1}}$. The median C-AGB $\mathrm{DPR}$ in the NGC~6822 bar is $8.4\times10^{-11}\,M_\odot\,\mathrm{yr^{-1}}$.

O-AGB stars and RSG stars produce metal oxides and silicate dust. The blue histogram in Figure~\ref{fig:dpr} represents the distribution of DPR values across these stellar classes. We again split this into two groups: a large dust-free population centred at $\log\mathrm{DPR}\approx -11.5$, which drops steeply, giving way to a second dust-producing group that peaks at $\log\mathrm{DPR}\approx -9.5$.
Most O-AGB stars in the sample have low-to-negligible dust-production rates, with 241/419 stars lying below the boundary of $\mathrm{DPR \leq 10^{-11.3}\,M_\odot\,yr^{-1}}$. These AGB stars are early in their evolution through the thermally pulsing stage, and their SEDs are characteristic of dust-free photospheres~\citep{DellAgli2016}; they may subsequently evolve into C-AGB stars after a third dredge-up event~\citep{Lattanzio2004}.

DPRs of order $\sim10^{-12}\,M_\odot\,\mathrm{yr^{-1}}$ correspond to effectively negligible dust production and reflect limitations in fitting photosphere-dominated SEDs \citep{Riebel2012, McDonald2025}. In this mass-loss regime, the inferred DPRs can be sensitive to assumptions such as reddening, which may bias their values toward higher DPRs. In this work, we correct only for foreground extinction towards NGC~6822 and do not include internal reddening. Since reddening within NGC~6822 varies spatially ($E(B-V) = 0.24$--$0.54$ mag; \citealt{Massey1995, HernandezMartinez2009}), this may introduce an additional uncertainty in the derived DPRs for these sources.

The total quantity of oxygen-rich dust produced in the sample is $\sum \mathrm{DPR}_{O}=3.4\times10^{-7}\,M_\odot\,\mathrm{yr^{-1}}$, with almost half of this ($1.6\times10^{-7}\,M_\odot\,\mathrm{yr^{-1}}$) coming from a single source.  Of the highest dust-producing stars (those with $\mathrm{DPR} \geq 10^{-8}\,M_\odot\,\mathrm{yr^{-1}}$), 6 out of 9 have oxygen-rich chemistry. 
These are likely higher-mass (4--8\,$M_\odot$) AGB stars undergoing HBB, consistent with spectroscopic confirmation of such sources in the bar of NGC~6822 \citep{Groenewegen2009}.
Massive O-AGB stars may begin producing dust within $\sim$30\,Myr of their formation~\citep{Boyer2017}, but they are observationally rare because of their short lifetimes~\citep{Justtanont2013,Whitelock2018}. Depending on the adopted gas-to-dust ratio, they may be losing mass at nearly $10^{-5}\,M_\odot\,\mathrm{yr^{-1}}$. Despite their small numbers, such massive O-AGB stars therefore make a disproportionate contribution to the dust returned to the ISM of NGC~6822.

RSG stars contribute $\mathrm{7.5\times10^{-9}\,M_\odot\,yr^{-1}}$ to the dust budget of NGC~6822, which amounts to around 1\% of the total DPR. However, this estimate is incomplete, as DPR measurements are unavailable for the brightest RSGs owing to saturation in the \jwst\ imaging, and the true RSG contribution may therefore be higher \citep{Antoniadis2025}.

Finally, most stars below the TRGB produce very little dust. Nevertheless, 106/363 lie above the DPR threshold and together contribute $\mathrm{2.4\times10^{-8}\,M_\odot\,yr^{-1}}$.  Modest mass loss on the RGB is typically described by Reimers' law~\citep{Reimers1975}, which parameterises the mass-loss rate as
\[
\dot{M} = 4\times10^{-13}\,\eta\,\frac{L R}{M},
\]
where $\eta$ is the scaling parameter (typically $\eta \sim 0.4$).  However, such winds are insufficient to account for the highest dust-production rates.

We find typical values of $\eta \sim 0.1$ with a dispersion of $\sim 0.5$ dex for our sample, indicating that while most TRGB stars are consistent with weak RGB winds, the highest dust-production rates would require significantly larger values of $\eta$ than are typical, and thus exceed what can be explained by the Reimers prescription.
RGB dust production has been observed above ${\sim}1000\,L_\odot$ \citep[$M_{\rm bol}=-2.8$;][]{McDonald2011}, but this is limited to the very upper tip of the RGB and remains small compared to AGB dust production~\citep{Groenewegen2014}.
The highest-DPR TRGB source produces $\mathrm{1.0\times10^{-8}\,M_\odot\,yr^{-1}}$ (42\% of the TRGB total), and 16 sources have $\mathrm{DPR}\geq10^{-10}\,M_\odot\,yr^{-1}$. The vast majority of TRGB sources (98\%) exhibit oxygen-rich chemistries, as expected for stars that have not yet undergone third dredge-up. These dust-producing objects may be misclassified RGB stars and instead represent TP-AGB stars observed mid–thermal pulse. To robustly distinguish these stellar classes, follow-up variability studies or spectroscopy would be required.

\subsection{Global Dust‑Production Rate of NGC~6822} \label{sec:res:Total_dpr}

We next consider how the measured DPR scales to the full extent of NGC~6822.
Our \jwst\ sample of AGB stars in NGC~6822 covers an area of approximately 29.0 square arcminutes. Although the stellar component of NGC~6822 extends to significantly larger radii \citep[deprojected radius ${\sim}1$–$2^\circ$;][]{Battinelli2006, Zhang2021}, the density of AGB stars declines steeply beyond the central bar \citep{Hirschauer2020}. The AGB-star catalogue produced by \citet{Hirschauer2020} spans a substantially larger footprint (3 square degrees) than the one studied here. Using \emph{UKIRT} and \spitzer\ imaging, they identified AGB candidates via a CMD kernel-density–based selection technique. Their catalogue contains 1610 reliable AGB stars, of which 458 lie within the NIRCam and MIRI fields of view used in our analysis (i.e.\ $\sim$28\% of the total). We therefore scale the total AGB dust-production rate measured within the \jwst\ footprint by this fractional coverage, assuming that the average dust-production rate per star does not vary strongly across the galaxy, obtaining a global AGB dust input of $\sim1.9\times10^{-6}\,\rm M_\odot\,yr^{-1}$. 
This conservative estimate does not correct for potential variations in stellar population ages, nor for completeness or crowding effects in the \citet{Hirschauer2020} catalogue. 

\begin{figure}
    \centering
    \includegraphics[width=\columnwidth]{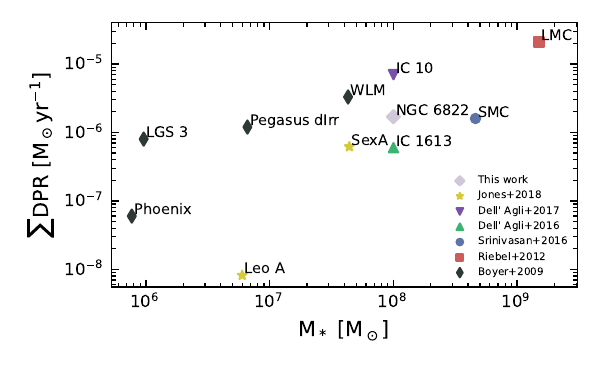}
	\caption[Cumulative DPR for Local Group Galaxies]{Global cumulative DPR from AGB stars as a function of stellar mass ($\rm M_*$) for Local Group galaxies. Stellar masses are from \citet{McConnachie2012}. DPR measurements are shown for WLM, Pegasus~dIrr, LGS~3 and Phoenix \citep[assuming a gas-to-dust ratio of 200;][]{Boyer2009}, IC~1613 \citep{DellAgli2016}, the LMC \citep{Riebel2012}, the SMC \citep{Srinivasan2016}, IC~10 \citep{DellAgli2017}, and Sextans~A and Leo~A \citep{Jones2018}.}
    \label{fig:total-dust}
\end{figure}

To place NGC~6822 in context within the Local Group, we show in Figure~\ref{fig:total-dust} the total DPR estimated for Local Group dwarf galaxies, versus their estimated stellar mass ($\rm M_*$), which are taken from \citet{McConnachie2012}. The DPR measured for NGC~6822 in this work lies comfortably within the range expected for galaxies of comparable stellar mass. In particular, the closest analogues are the SMC \citep[$1.6\times10^{-6}\,\rm M_\odot\,yr^{-1}$;][]{Srinivasan2016}, IC~1613 \citep[$6\times10^{-7}\,\rm M_\odot\,yr^{-1}$;][]{DellAgli2016}, and IC~10 \citep[$7\times10^{-6}\,\rm M_\odot\,yr^{-1}$;][]{DellAgli2017}.
The metallicities of these systems provide a natural explanation for the observed spread in DPRs. IC~10 \citep[${\sim}0.4Z_\odot$;][]{Garnett1990,Magrini2009}, being the most metal‑rich of the three, is expected to exhibit enhanced dust production owing to the greater efficiency of dust formation at higher $Z$ \citep{Hofner2018}. 
The SMC, whose metallicity is similar to that of NGC~6822 \citep[${\sim}0.2Z_\odot$;][]{Garnett1995,Peimbert2000}, displays a dust‑production rate closely matching our measurement. NGC~6822’s dust production therefore fits naturally within the metallicity‑driven trends observed in other Local Group dwarfs.

AGB dust production in low‑metallicity galaxies -- such as IC~1613, the SMC, and the slightly more metal‑rich LMC \citep{Riebel2012} -- is typically found to be dominated by C-AGB stars. In NGC~6822, however, our results indicate the opposite trend: roughly two‑thirds of the dust originates from O-AGB stars. The recent star‑formation history of NGC~6822 may explain this disparity. The galaxy experienced enhanced star formation within the past 100–200 Myr \citep{Gallart1996b}, which, at the metallicity of NGC~6822, would produce oxygen-rich AGB stars with initial masses of $\sim4\,M_\odot$, capable of undergoing hot-bottom burning \citep[][]{Marigo2008}. Such stars are efficient dust producers, and their presence could account for the elevated oxygen‑rich DPR measured in our sample. 

Misclassification of some of the most extreme sources offers another possible explanation for the elevated oxygen‑rich dust contribution.  The highest dust-producing object in the sample, CN123349, shown in the lower-right panel of Figure~\ref{fig:res:sed} is fitted equally well by AGB and YSO SED models and lies just below the adopted boundary in Figure~\ref{fig:res:cleaning}. Given the uncertainties inherent in SED fitting, it is not possible to determine whether this source is truly producing $\rm 1.6\times10^{-7}\,M_\odot\,yr^{-1}$ of oxygen‑rich dust, or whether it is instead a deeply embedded YSO, which would remove rather than contribute dust to the ISM. 
Excluding this source yields nearly equal cumulative carbon‑ and oxygen‑rich dust contributions, although the oxygen‑rich dust contribution remains higher than in other metal‑poor galaxies.  Definitive classification of these extreme sources will require follow‑up spectroscopy or narrow‑band photometry. The contribution to the total dust production by the remaining 13 ``ambiguous objects''  (i.e.\ the $\Delta{\rm AIC} \leq 0.5$ sources from Section~\ref{sec:res:clean}, excluding CN123349) is modest in comparison.

\subsection{Observable Proxies for Dust Production}
\label{sec:res:dpr_relation}

Given the large number of photometric bands available with \jwst, it is useful to identify simple colour‑based diagnostics that can serve as empirical proxies for AGB dust‑production rates. Such empirical relations have long been established using \emph{IRAS} data \citep{Whitelock1994,LeBertre1997}. In particular, the $\mathrm{K-[12]}$ versus DPR relation is notably tight because the near‑IR filter traces the stellar photosphere while the mid‑IR filter probes emission from the circumstellar envelope. This combination therefore provides a simple and effective observable proxy for estimating dust‑production rates. 
Similar correlations have been established using \spitzer\ photometry \citep{Matsuura2009,Gullieuszik2012,Riebel2012}, where DPR shows clear trends with $\mathrm{K-[8.0]}$ and $\mathrm{[3.6]-[8.0]}$ colours. 
We therefore examine whether comparable empirical relations can be established using \jwst\ filter combinations.

\begin{table}
    \centering
    \caption[Observational proxy for Dust Production Rate]{Best‑fitting parametrised hyperbolic functions for estimating C-AGB dust‑production rates from \jwst\ NIRCam–MIRI colour indices. Each relation adopts the form $\mathrm{DPR}=p_0/(x+p_1) + p_2$.  Listed are the fitted parameters and the median absolute deviation (\textsc{mad}) of each colour combination.}
    \begin{tabular}{l|cccc}
    \hline
    \hline
        \jwst\ Colour &$p_0$ &$p_1$ &$p_2$ &\textsc{mad} \\
        \hline
        F115W--F1500W &  -28.20 &   4.21&   -6.21&    0.28\\
        F115W--F2100W &  -37.84 &   5.72&   -5.81&    0.19\\
        F200W--F770W  &  -12.62 &   2.46&   -6.68&    0.23\\
        F200W--F1000W &   -5.91 &   1.26&   -7.68&    0.09\\
        F200W--F1500W &  -15.82 &   2.89&   -6.45&    0.19\\
        F200W--F2100W &  -16.52 &   2.98&   -6.45&    0.13\\
        F356W--F770W  &   -0.65 &   0.04&   -9.23&    0.27\\
        \hline
    \end{tabular}
    \label{tab:proxies}
\end{table}

\begin{figure}
    \centering
    \includegraphics[width=\columnwidth]{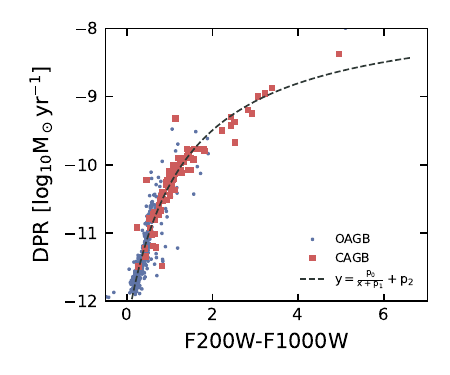}
    \caption[Observational proxies to measure dust-production rate]{$\mathrm{DPR}$ versus F200W--F1000W colour for C-AGB (red) and O-AGB (blue) stars.  The best‑fitting hyperbolic function,  $\mathrm{DPR}=p_0/(x+p_1)+p_2$ is shown as a dashed line.}
    \label{fig:proxy}
\end{figure}

Figure~\ref{fig:proxy} shows the relation between F200W--F1000W colour and $\mathrm{DPR}$ for C-AGB and O-AGB stars.  We fit a hyperbolic function of the form $\mathrm{DPR}=p_0/(x+p_1)+p_2$, where $x$ denotes the colour index and $p_0$, $p_1$, and $p_2$ are free parameters. This parametrisation provides a good description of the C-AGB population, whose infrared colours span a broad range, in contrast to the more tightly clustered colours of O-AGB stars. We do not fit an equivalent relation for O-AGB stars because their infrared colours span too narrow a range to constrain a meaningful function.
To evaluate the quality of each fit, we compute the median absolute deviation of the residuals (\textsc{mad}) and list all colour combinations with $\textsc{mad}\leq0.3$ in Table~\ref{tab:proxies}. Achieving such low‑scatter relations requires wide colour baselines, and therefore at least one NIRCam and one MIRI filter. This reflects the increasing influence of circumstellar dust on the SED at longer wavelengths. By sampling flux in a shorter NIRCam filter, which primarily traces the stellar photosphere, together with a dust‑sensitive MIRI filter, we can effectively measure the relative strengths of the photospheric and circumstellar components.

We find that colour combinations involving the F200W NIRCam filter yield the lowest \textsc{mad} values and therefore the tightest empirical relations when paired with any of the available MIRI bands. Although slightly bluer than the classical $K$ band, F200W samples a similar region of the stellar SED. When combined with F1000W, which probes the rising edge of the 11.3\,$\mu$m SiC feature in carbon‑rich circumstellar shells, the resulting colour index provides a natural analogue to the $\mathrm{K-[12]}$ diagnostic. At longer MIRI wavelengths, dust emission becomes increasingly prominent, and both F200W--F1500W and F200W--F2100W produce well‑constrained fits. However, the source density drops sharply in F2100W, with only 135 AGB stars detected in both bands, making the calibration of these relations more uncertain.

\subsection{Spatial Distribution of Dust Production}

\begin{figure}
	\centering
	\includegraphics[width=\columnwidth]{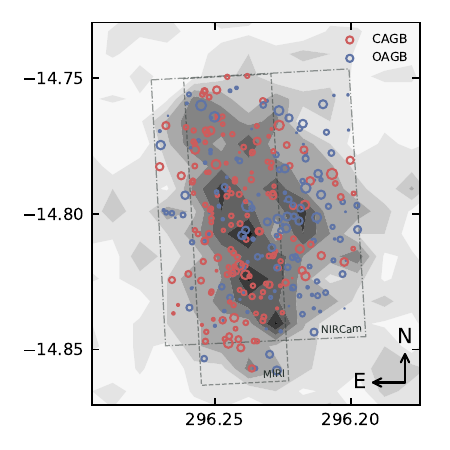}
	\caption[Spatial dust-production distribution]{Spatial distribution of dust-producing (DPR$\geq10^{-11.3}$) AGB stars, coloured by chemical subtype (O-AGB: blue; C-AGB: red), with point size scaled by individual DPR. The underlying contour map shows the AGB population from \citet{Hirschauer2020}. The NIRCam and MIRI fields of view are indicated by dashed boxes.}
	\label{fig:globaldust}
\end{figure}

Figure~\ref{fig:globaldust} shows the spatial distribution of the dusty AGB population across the NGC~6822 \jwst\ fields. For illustrative purposes, the size of each point scales with the DPR. The contour map traces the full AGB population identified by \citet{Hirschauer2020} from UKIRT near-infrared \citep{Sibbons2012} and \textit{Spitzer} mid-infrared \citep{Khan2015} photometry. The 14.5\,arcmin$^2$ MIRI and 29.0\,arcmin$^2$ NIRCam footprints are outlined by dashed boxes. The dust‑producing AGB stars broadly trace the underlying AGB distribution identified by earlier near‑IR and mid‑IR surveys. A clear chemical segregation is, however, apparent: the central bar contains a significantly higher fraction of carbon‑rich stars, whereas oxygen‑rich stars are more spatially extended. Previous studies \citep{Sibbons2012,Hirschauer2020} have similarly reported an increasing O-AGB:C-AGB ratio with radial distance in NGC~6822, although small‑number statistics have limited the robustness of those measurements.

Intermediate‑age AGB stars in dwarf galaxies are not expected to migrate far from their birth sites \citep[e.g.][]{El-Badry2016}, so local variations in the distributions of O-AGB and C-AGB stars may preserve signatures of past star‑formation episodes. O-AGB stars can arise from both older and intermediate‑age populations, whereas C-AGB stars predominantly trace intermediate‑age stars that have undergone third dredge‑up. Spatial differences between the two classes may therefore provide insight into the star‑formation history of NGC~6822 in addition to its metallicity structure. We note that classification uncertainties are larger in NIRCam‑only regions, where the absence of mid‑IR photometry reduces sensitivity to circumstellar dust features, making chemical typing less secure than in areas with full NIRCam–MIRI coverage.

Mapping the ratio of carbon‑rich to oxygen‑rich AGB stars (the C/M ratio) is a long‑established method for tracing metallicity variations across galaxies \citep{Cioni2003}. The surface oxygen abundance of O-AGB stars is set primarily by their initial metallicity, whereas C-AGB stars become carbon‑rich through the addition of carbon to the envelope during third dredge‑up. In metal‑poor environments, the initially low oxygen abundance means that relatively modest carbon enrichment is sufficient to raise the surface C/O ratio above unity \citep{Iben1983,Battinelli2005}, increasing the number of C-AGB stars relative to O-AGB stars. The resulting C/M ratio, therefore provides an effective empirical proxy for metallicity. 

\begin{figure*}
	\centering
	\includegraphics[width=\textwidth]{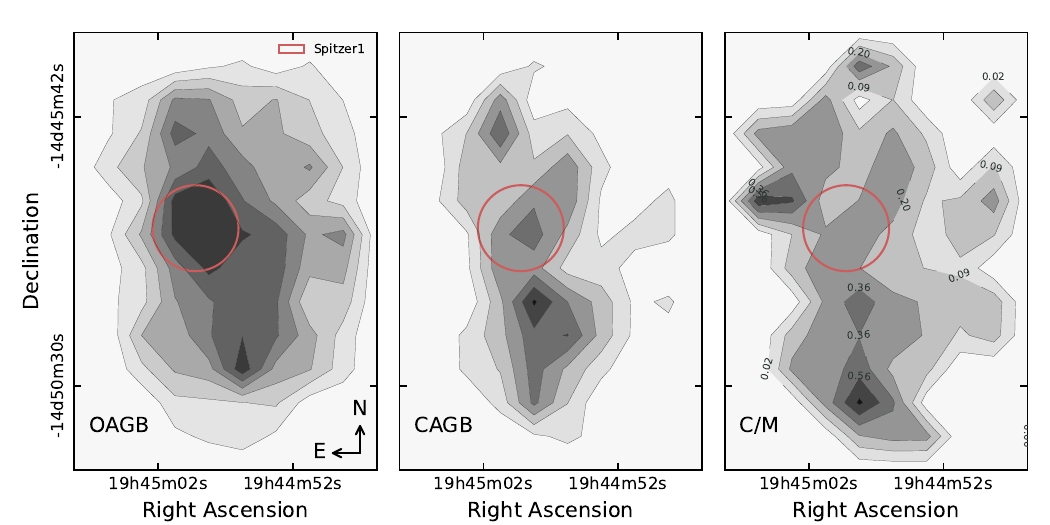}
	\caption[C/M ratio maps for the NGC~6822 central bar]{Source‑density contour maps for O-AGB stars (left), C-AGB stars (middle), and the C/M ratio (right). Sources are binned in 0.6\arcmin\ cells and smoothed with a 0.24\arcmin\ Gaussian kernel. The location of Spitzer~I, a prominent star‑forming region, is indicated in red.}
	\label{fig:comap}
\end{figure*}

We construct density maps by counting O-AGB and C-AGB stars in 0.6\arcmin\ bins in right ascension and declination. These maps are then smoothed with a 0.24\arcmin\ Gaussian kernel and displayed as contour plots in Figure~\ref{fig:comap}. The left‑hand panel shows the distribution of O-AGB stars, while the middle panel shows C-AGB stars. The right‑hand panel presents the C-AGB/O-AGB density ratio, smoothed in the same manner. The location of Spitzer~I \citep{Jones2019,Lenkic2024}, a prominent massive star‑forming region, is marked in red.

The C/M ratio in NGC~6822 ranges from $\sim1$ on global scales (including the galaxy’s extended stellar halo) to $\lesssim0.3$ in the central bar \citep[e.g.][]{Letarte2002,Cioni2005,Sibbons2012,Sibbons2015,Kacharov2012,Kang2006}. \citet{Kang2006} report a C/M ratio of $0.27\pm0.03$ in the central bar, with a gradient from $\sim0.22$ in the north to $\sim0.31$ in the south. Our \jwst\ measurements span a C/M $\sim$0.2--0.36 across the bar, with broadly consistent spatial trends (Figure~\ref{fig:comap}). We also identify a localised peak reaching $\sim0.5$, although this is confined to a small area and may reflect small-number statistics.  Our total counts include 419 O-rich and 191 C-rich AGB stars, yielding C/M $\sim0.45$. Including red supergiants candidates in the M-type population reduces this to $\sim0.28$.

The density of O-AGB stars is elevated in the region overlapping Spitzer~I, which may indicate residual YSO contamination. However, the AGB stars in this region are almost entirely dust‑free, and their SEDs are consistent with photospheric emission rather than the dust‑dominated SEDs typical of early‑stage YSOs \citep{Jones2017a}. 
This suggests that the AGB‑sample cleaning method described in Section~\ref{sec:res:clean} is effective at identifying dust‑free AGB stars. Furthermore, the fraction of O-AGB stars with a dust excess, those most likely to be confused with YSOs, is not enhanced in this region compared with the rest of the field, indicating that any remaining contamination is minimal.
Localised peaks in the C/M ratio map are likely driven by small‑number statistics, and the limited spatial coverage of our \jwst\ fields restricts its use as a global metallicity diagnostic. Even so, the spatial distribution of dust‑producing AGB stars and their chemical subtypes reveals a coherent structure linked to both star‑formation history and circumstellar dust production.  Wide‑field \euclid\  \citep{Hunt2025} and future \hst medium-band imaging \citep{Nally2025_HST, Jones2025_HST}, combined with the \jwst\ results presented here, will enable a comprehensive reconstruction of the C/M ratio and AGB dust‑production patterns across the entire galaxy.

\section{Summary and Conclusions}\label{sec:conclusions}

We have constructed a deep \jwst\ NIRCam and MIRI point‑source catalogue of NGC~6822 using PSF photometry and band‑matching with \starbug. The final catalogue contains 864{,}114 unique NIRCam detections, 17{,}235 MIRI detections, and 10{,}079 sources detected in both instruments. Blackbody fitting yields effective temperatures and bolometric luminosities for 119{,}621 stars, reaching a depth of ${\sim}$0.5\,mag below the red clump and enabling a detailed characterisation of the galaxy’s resolved stellar populations. The full photometric catalogue and all derived parameters are provided in machine‑readable form.

Using these data, evolved‑star candidates were selected from colour–magnitude diagrams, with resolved contaminants removed via visual inspection and young stellar objects excluded through SED fitting with the \citet{Robitaille2017} models. In total, 1226 evolved‑star candidates were retained and classified as carbon‑rich AGB, oxygen‑rich AGB, RSG, or TRGB stars using {\sc grams} radiative‑transfer models,  which also return dust‑production rates for each source. 

The carbon‑star luminosity function peaks at $M_{\rm bol}\approx -4.5$, consistent with expectations for a low‑metallicity system, and we identify a small population of unusually luminous AGB candidates that may represent super‑AGB stars.
Across the {\em JWST} fields, we measure a total dust‑production rate of $5.6\times10^{-7}\, M_\odot\,{\rm yr^{-1}}$, dominated by O‑rich AGB stars (60\%), with C‑rich AGB stars contributing 35\% and the remainder arising from RSG and TRGB sources.  By combining our results with AGB number counts from the wider, lower‑resolution survey of \citet{Hirschauer2020}, we infer a conservative global dust-production estimate of $1.9\times10^{-6}\, M_\odot\,{\rm yr^{-1}}$, consistent with Local Group galaxies of similar mass and metallicity.
The high fractional contribution of O‑rich AGB stars to the dust budget is atypical for a low‑metallicity system. It indicates the presence of intermediate‑mass O‑rich AGB stars undergoing hot‑bottom burning, consistent with NGC~6822’s recent star‑formation history.

Spatially, the dusty AGB population exhibits a centrally concentrated carbon‑rich component and a more extended oxygen‑rich component, reflecting the known stellar‑population gradients across the galaxy’s bar. 
We also derive empirical {\em JWST}  colour relations that provide effective photometric proxies for dust‑production rates, particularly when combining one NIRCam and one MIRI filter. These results provide the most detailed census to date of evolved stars and dust production in NGC~6822 and establish a foundation for future wide‑field mapping and spectroscopic studies aimed at tracing mass loss, dust chemistry, and the chemical evolution of metal‑poor dwarf galaxies.

\

\noindent {\it Facilities:} {\em JWST} (NIRCam \& MIRI) - James Webb Space Telescope.

\noindent {\it Software:} {\sc astropy} \citep{Astropy2013}, {\sc photutils} \citep{photutils}, {\sc scipy} \citep{scipy}, \starbug\ \citep{starbug2} , and {\sc topcat} \citep{Taylor2005}.

\section*{Acknowledgements}

This work is based on observations made with the NASA/ESA/CSA James Webb Space Telescope. The data were obtained from the Mikulski Archive for Space Telescopes at the Space Telescope Science Institute, which is operated by the Association of Universities for Research in Astronomy, Inc., under NASA contract NAS 5-03127 for {\em JWST}. These observations are associated with program \#1234.
CN acknowledge the support of an STFC studentship (2645535).
OCJ has received funding from an STFC Webb fellowship.
MM and NH acknowledge support through NASA/JWST grant 80NSSC22K0025, and MM and LL acknowledge support from the NSF through grant 2054178.
MM, NH, and LL acknowledge that a portion of their research was carried out at the Jet Propulsion Laboratory, California Institute of Technology, under a contract with the National Aeronautics and Space Administration (80NM0018D0004).
AMNF is supported by UK Research and Innovation (UKRI) under the UK government’s Horizon Europe funding guarantee [grant number EP/Z534353/1] and by the UK Science and Technology Facilities Council [grant number ST/Y001281/1].
PJK acknowledge support from Research Ireland under Grant Number 21/PATH-S/9360.

\section*{Data Availability}

The data used in this study may be obtained from the Mikulski Archive for Space Telescopes (MAST; \href{https://mast.stsci.edu/}{https://mast.stsci.edu/}) and are associated with program \#1234.



\bibliographystyle{mnras}
\bibliography{dust} 




\bsp	
\label{lastpage}
\end{document}